\documentclass[superscriptaddress,prb,super=false,
showpacs,floatfix,fleqn,longbibliography]{revtex4-1}
\usepackage[dvips]{graphicx}
\usepackage{epsfig}
\usepackage{subfig}
\usepackage{color}
\usepackage[normalem]{ulem}
\usepackage{amsmath,amsfonts,amssymb,bm}%,ulem}
\usepackage{multirow}
\usepackage[table]{xcolor}
\UseRawInputEncoding
\setcitestyle{numbers,square}

\begin{document}
\setlength{\tabcolsep}{.5em}
\renewcommand{\arraystretch}{1.2}

\title{The origin of anomalous non-linear microwave absorption in Josephson junction qubits: mysterious nature of two level systems  or their dynamic interaction?}
\author{Alexander L. Burin}
\affiliation{Tulane University, New
Orleans, LA 70118, USA}

\date{\today}
\begin{abstract}
Quantum two-level systems (TLSs) commonly found at low temperature in amorphous and disordered materials are  responsible for decoherence in superconducting Josephson junction qubits particularly because they absorb  energy of coherent qubit oscillations in the microwave frequency range.  In planar Josephson resonators with oxide interfaces this absorption is characterized by an   anomalously weak  loss tangent dependence on the  field in the non-linear regime that conflicts  with the theoretical expectations and   the observations in amorphous dielectrics.   It was recently  suggested that this anomalous absorption is due to  TLS dynamic interactions. Here we  show  that  such   interactions cannot lead to  the observed loss-tangent field dependence  and suggest the alternative explanation  assuming that TLS dipole moments $p$ are distributed according to the specific power law $P(p) \propto 1/p^{3-\eta}$ ($0\leq \eta <1$).   This assumption, indeed, results in the observed loss tangent behavior.  The hypothesis of a power law distribution  is  supported both by the recent measurements  of individual TLS dipole moments  and the theoretical model of TLS formation due to the long-range dipole-dipole interaction,  thus connecting  the anomalous absorption with  the possible solution of the long-standing problem of the nature of TLSs. 
\end{abstract}

\maketitle
%\small{
%\yg{We still have space. I believe that the following remains
%  to be done:
%\begin{itemize}
%\item Agree on numerical estimates.
%\end{itemize}
%}

%\section{Introduction}
%\label{sec:intro}

\section{Introduction} 
\label{sec:Intr}

Quantum two level systems (TLSs) substantially limit performance of superconducting and optomechanical  resonators \cite{Cao2007NoiseInSCRes,Park2009OptMech} and  superconducting qubits \cite{Martinis05,Ustinov2012ScienceTLS}, which are one of the most promising candidates for use in quantum hardware  \cite{Martinis05,Yu2004,Lisenfeld2019ReviewTLSs,
McDermott2009ReviewSuperQub,oliver_welander_2013RevQub,
KlimovMartinis2018FluctEnergRelax,Bilmes2021TLSNat}. The existence of TLSs in practically all disordered materials and their quantitative universality  \cite{PhillipsReview,Hunklinger1986265Rev,YuLeggett88,Enss02Rev}, make their destructive effect seemingly unavoidable.  Observation  and interpretation of  any deviation from that universal behavior can  shed  light on a long-standing, yet unresolved problem of the nature of TLSs and, based on this knowledge, suggest the way to reduce their deleterious effect.  Here we focus on the anomalous absorption of microwaves in certain Josephson resonators as such deviation and show that  the observed  anomalous field  dependence of absorption  naturally emerges if TLSs   are formed as  universal  low energy collective excitations in a system with  the long-range interaction  \cite{YuLeggett88,ab96RG,ab98book}.  Our theory  is different from   the earlier suggested explanation of anomalous absorption using TLS dynamic interactions \cite{Faoro12TLSInt}.  To address this conflict we also consider  the absorption in the presence of a dynamic  interaction and show that any such interaction  cannot lead to  the observed behavior.  

%figAbsPr.eps

\begin{figure}[h!]
%\centering
\includegraphics[width=\columnwidth]{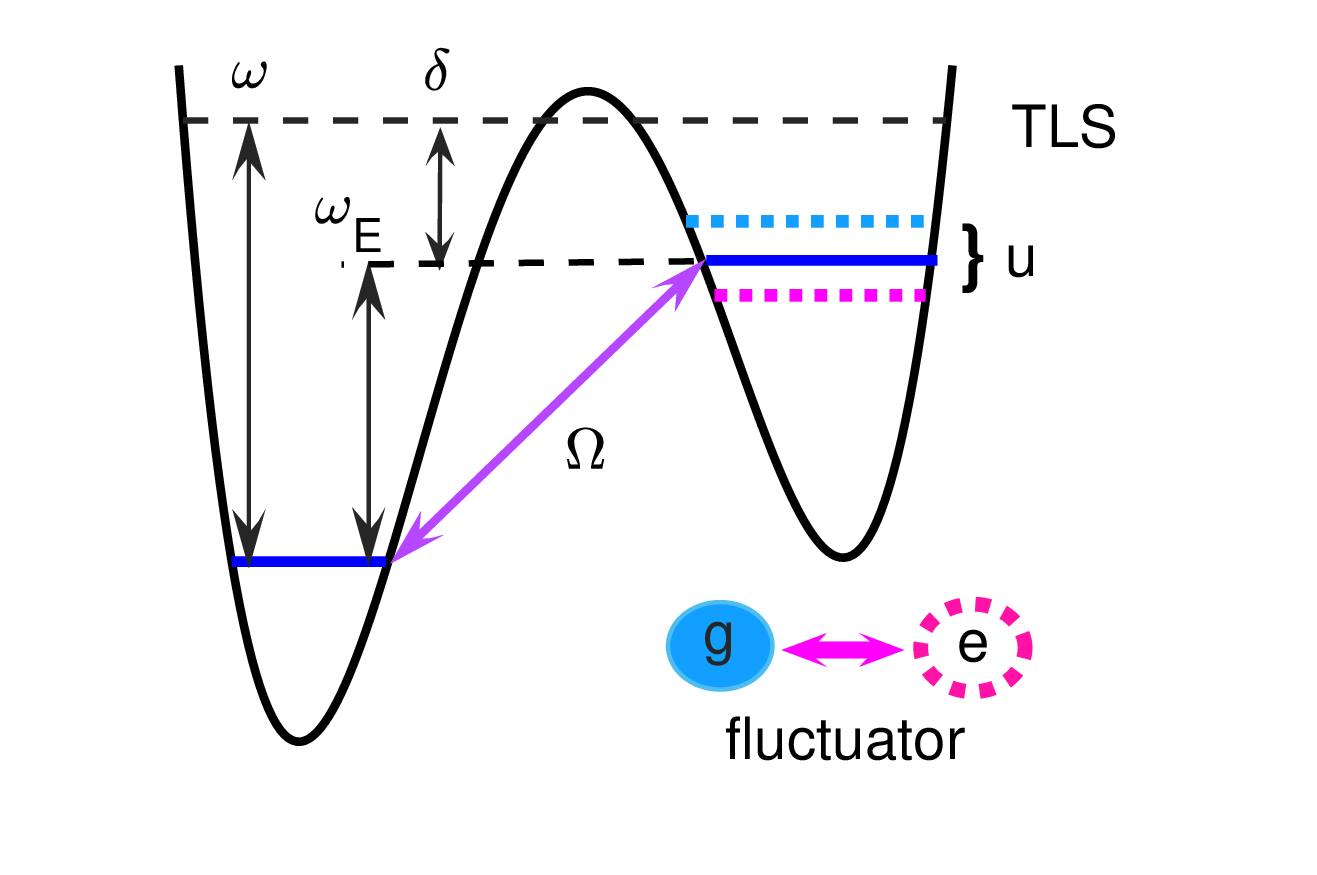}
\caption{\small TLS  is represented by two lowest energy levels (solid lines) for atom or group of atoms  	positioned within a double  well potential \cite{AHV,Ph}.      TLS interacts with the microwave field  having  the frequency $\omega$ shifted from the TLS resonant  frequency  $\omega_{E}=E/\hbar$ ($E$ is TLS energy splitting) by detuning $\delta$.   This field  couples two TLS states with the strength determined by the Rabi frequency $\Omega$.   Two state fluctuator shown in the bottom right corner can be in its   ground (g) or excited (e) states represented by filled and empty circles,  respectively.  Fluctuator transitions  between these states shift  TLS resonant frequency by $\pm u/2$,  as shown by dotted lines,  where $\hbar u$ is the  TLS interaction with the fluctuator. }
\label{fig:TLS}
\end{figure}

% To describe  the anomalous absorption behavior and the previous theoretical attempt to interpret it \cite{Faoro12TLSInt} based ion TLS interaction with some other defects called fluctuators,  we introduce the readers to the problem in necessary detail.  
  
 Before considering the anomalous nonlinear absorption   we briefly introduce   present knowledge   in the area  of resonant microwave  absorption  by TLSs.   TLSs are commonly represented  by two lowest energy levels in double well potentials (see Fig.  \ref{fig:TLS})  having  energy splitting $E$ (shown in  Fig.  \ref{fig:TLS} in frequency units as $\omega_{E}=E/\hbar$) and coupled by the tunneling characterized by the amplitude $\Delta_{0}$ (not shown). They  obey the universal probability density function defined as  \cite{AHV,Ph,PhillipsReview,Hunklinger1986265Rev} 
 \begin{eqnarray}
 P(E, \Delta_{0})=P_{0}\frac{E}{\Delta_{0}\sqrt{E^2-\Delta_{0}^2}}. 
 \label{eq:AHWP}
 \end{eqnarray}
Since only TLSs with large tunneling amplitudes $\Delta_{0}\sim E$ contribute to the resonant absorption \cite{Hunklinger1986265Rev}) we restrict our qualitative consideration to TLSs with $\Delta_{0} \sim E$ and use Eq. (\ref{eq:AHWP}) as the uniform distribution over TLS energies and in space in the form $P(E) \approx P_{0}$.  
 
TLS interacts with the external microwave field characterized by the frequency $\omega$ and amplitude $\mathbf{F}$,  because tunneling atoms possess electric charges,  so each TLS possesses a  dipole moment $\mathbf{p}$.  Resonant TLS ($E\approx \hbar\omega$) can transfer between its levels due to interaction with the field characterized by  the Rabi frequency  $\Omega$, where $\hbar\Omega \approx  \mathbf{Fp}\Delta_{0}/E$  \cite{ab13LZTh}. Since for efficiently absorbing  TLSs  one has $\Delta_{0} \sim E$, we use the maximum Rabi frequency $\Omega =pF/\hbar$ as its typical value assuming all TLS dipole moments approximately identical.   

Deviation of TLS resonant frequency from the field frequency  is determined   by its frequency detuning $\delta=\omega-\omega_{E} \ll \omega$ ($\omega_{E}=E/\hbar$, see Fig.  \ref{fig:TLS}).     Absorption can be affected by neighboring dynamic fluctuators (e. g. other TLSs) changing  TLS detuning   \cite{Faoro12TLSInt} when transferring between their states.  One such fluctuator modifying TLS detuning by $\pm u/2$  is shown in  Fig. \ref{fig:TLS}. 
% The fluctuator, i. e.  another two level defect  (see Fig. \ref{fig:TLS}),  affects TLS detuning shifting it by $\pm u/2$ in its two different states.  Here $u$ stands for the TLS-fluctuator interaction expressed in the frequency units.   %We introduced fluctuators in our consideration because they were used in Ref. \cite{Faoro12TLSInt} to 
 
 Energy absorption by TLSs emerges due to their interaction with the environment, where the absorbed energy is dissipated.   We represent the  TLS interaction with the environment  in terms of its  relaxation and decoherence times  $T_{1}$ and $T_{2}$, respectively  \cite{VONSCHICKFUS1977144,Hunklinger1986265Rev},  similarly to that for nuclear   spins \cite{Bloembergen1948spindec,EssentialNMRBook}.   Since efficiently absorbing  TLSs possess tunneling amplitudes comparable with their energies, the times $T_{1}$ and $T_{2}$ express minimum relaxation and decoherence times for TLSs with energy of order of $\hbar\omega$ \cite{Hunklinger1986265Rev}.  
 
 Consider resonant absorption   of microwaves by TLSs at low temperatures  $k_{B}T \ll \hbar\omega$ which is true  in most of experiments \cite{Martinis05,Lisenfeld2019ReviewTLSs,Kevin2023PowWeakDep} and   in the absence of fluctuators.   We characterize  TLS absorption in terms of its loss tangent  (inverse quality factor)  represented by  the  energy absorption rate per a unit volume divided by  the product of the microwave energy density $U_{F}=\epsilon \epsilon_{0}F^2$ and  the microwave field frequency $\omega$.  This parameter expresses  the ratio of the imaginary part of TLS contribution to the dielectric constant and the real part of the dielectric constant. 
 
\begin{figure}[h!]
\centering
\includegraphics[width=\columnwidth]{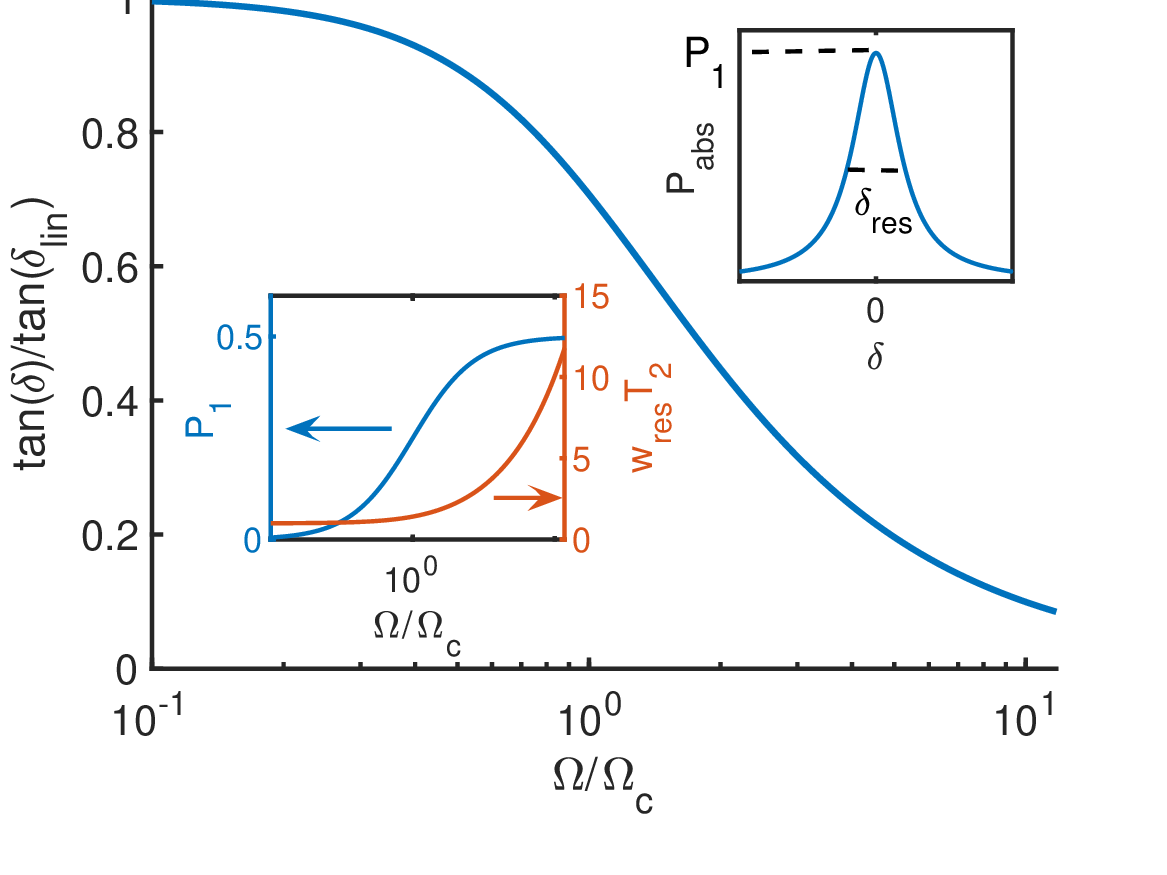}
\caption{\small Dependence of loss tangent on noninearity parameter $\Omega/\Omega_{c}$ Eq. (\ref{eq:lincr}). 
Loss tangent is proportional to the product of the absorption probability $P_{1}$ at resonance and the width of resonance $\delta_{\rm res}$ divided by the intensity proportional to $\Omega^2$.   Dependencies of $P_{1}$ and $\delta_{\rm res}$ on the nonlinearity parameter expressed as a ratio $\Omega/\Omega_{c}$  are shown  in the bottom left inset, while the dependence of the absorption probability $P_{1}$ on detuning $\delta$ is shown in the top right inset. The resulting dependence of the loss tangent ($\tan(\delta) \propto P_{1}\delta_{\rm res}/F^2$) on the non-linearity parameter is shown in the main graph.}
\label{fig:Abs}
\end{figure}

 In the absence of microwaves TLS occupies its ground state.    Microwave  field results in  its gradual  excitation   interrupted eventually by this  TLS relaxation back to its ground state  occurring over  the time of order of its relaxation time $T_{1}$.  The average absorbed energy during the relaxation time is given by the product of the probability $P_{1}$ of TLS transfer to  the excited state  during the time $T_{1}$ and the absorbed energy $E\approx \hbar\omega$. The energy absorption rate by a single TLS can be then approximated by $W_{\rm TLS} \approx \hbar\omega P_{1}/T_{1}$.   The probability $P_{1}$ depends on TLS detuning $\delta$ and it decreases with increasing $|\delta |$ as illustrated in  the top right inset in Fig. \ref{fig:Abs}.    We characterize the absorption  by the width  of the  absorption resonance  $\delta_{\rm res}$, which can be defined as $P_{1}(\delta_{\rm res})\approx P_{1}(0)/2$  (see  the top right inset in Fig. \ref{fig:Abs}). The number of absorbing TLSs per a unit volume  is determined by the probability density  Eq.  (\ref{eq:AHWP}) as $N_{\rm TLS}\approx P_{0}\hbar\delta_{\rm res}$.   Consequently, we estimate  the TLS loss tangent expressing the TLS  energy absorption rate as the product of the individual TLS absorption rate $W_{\rm TLS}$ and the number of absorbing TLSs $N_{\rm TLS}$.   This yields 
 \begin{eqnarray}
 \tan(\delta)\approx \frac{P_{0}\hbar^2\delta_{\rm res}P_{1}(0)}{T_{1}U_{F}},   ~ U_{F}=\epsilon \epsilon_{0}F^2.  
 \label{eq:AbsLT}
 \end{eqnarray} 
 
 For $T_{2} \ll T_{1}$ and at  small fields where the absorption probability is small ($P_{1}(0) \ll 1$), this probability is given by the sum of  $n=T_{1}/T_{2}$ probabilities $P_{2}$ of absorption in $n$ phase uncorrelated time domains of the size $T_{2}$.   The probability of absorption during the decoherence time $T_{2}$ by TLS having  zero detuning is given  by the quantum mechanical expression $P_{2} \approx \Omega^2T_{2}^2$ and, consequently,  $P_{1}(0) \approx \Omega^2T_{1}T_{2}$, which is valid only for $P_{1}(0) \ll 1$. Consequently, the probability $P_{1}$ is  small under the condition \cite{VONSCHICKFUS1977144,Hunklinger1986265Rev,PhillipsReview}   
\begin{eqnarray}
\Omega < \Omega_{c}= \frac{1}{\sqrt{T_{1}T_{2}}}. 
\label{eq:lincr}
\end{eqnarray}
 The probability estimate $P_{1}(0)$ is valid for detuning less than the inverse decoherence time, while at larger detuning $\delta> 1/T_{2}$ one has $P_{2} \approx  \Omega^2/\delta^2$ and  $P_{1} \approx  \Omega^2T_{1}/(\delta^2T_{2})$.  Consequently,  $\delta_{\rm res} \approx 1/T_{2}$ (see Fig. \ref{fig:Abs}) and we estimate the loss tangent using Eq. (\ref{eq:AbsLT}) as $\tan(\delta) \approx P_{0}p^2/(\epsilon_{0}\epsilon)$.  Thus at small fields satisfying Eq. (\ref{eq:lincr}) the loss tangent is field independent and approaches its maximum emerging in the linear response regime \cite{ab13LZTh}. 
 
 At larger fields the absorption saturates at $\delta < \delta_{\rm res} \approx \Omega\sqrt{T_{1}/T_{2}}$ where the probability   $P_{1}(\delta_{\rm res}) \approx \Omega^2T_{1}/(\delta^2T_{2})$ approaches unity  (see  Fig. \ref{fig:Abs}),   In this {\it nonlinear regime} the loss tangent evaluated using Eq. (\ref{eq:AbsLT})  acquires an  inverse field dependence in the form  $\tan(\delta_{\rm nl}) \approx \hbar^2P_{0}\delta_{res}/(T_{1}U_{F})\approx  P_{0}p^2/(\epsilon_{0}\epsilon\Omega \sqrt{T_{1}T_{2}}) \propto 1/F$ and the nonlinear regime takes place at $\Omega>\Omega_{c} \approx 1/T_{1}$.

 % define Delta_0 and true distribution  loss tangent  approx/exact
 
Our estimates for the loss tangent are very close to the results of the accurate calculations \cite{VONSCHICKFUS1977144,Martinis05,ab13LZTh} for the linear ($\tan(\delta_{\rm lin})$) and  nonlinear ($\tan(\delta_{\rm nl})$) regimes  
  \begin{eqnarray}
\tan(\delta_{\rm lin}) =\frac{\pi}{3}\frac{P_{0}p^2}{\epsilon\epsilon_{0}},  ~\delta_{\rm res}\approx \frac{1}{T_{2}}, ~ \Omega <  \frac{1}{\sqrt{T_{1}T_{2}}},
\nonumber\\
\tan(\delta_{\rm nl}) =\frac{3\pi}{8}\frac{\tan(\delta_{\rm lin})}{\Omega \sqrt{T_{1}T_{2}}},  ~ \delta_{res}\approx \Omega \sqrt{\frac{T_{1}}{T_{2}}},  ~  \Omega >  \frac{1}{\sqrt{T_{1}T_{2}}}.  
\label{eq:tand0}
\end{eqnarray}
At low temperatures where the TLS-TLS interaction is negligibly small, decoherence is associated with the relaxation and one has $T_{2}=2T_{1}$ similarly to the magnetic resonance problem  \cite{Hunklinger1986265Rev,ab13LZTh}.  In this case the size of a resonant domain in a linear regime   is $\delta_{\rm res}\approx 1/T_{1}$, while in the nonlinear regime it is approximately equal to the Rabi frequency $\Omega$. 
 
The loss tangent behavior Eq.  (\ref{eq:tand0}) was found consistent with experiments in bulk amorphous materials \cite{Hunklinger1986265Rev}.   However,  there is a common observation  \cite{Siegel2010LossWeakPowDepAl,Osborn2011NonlinWeakPowDep,Welander2011TLSAbsNLin,Faoro12TLSInt,
ab17Katz,Kevin2023PowWeakDep} in planar Josephson resonators with oxide interfaces  that TLS dielectric losses  does not follow that  theoretical prediction and TLS loss tangent  depends on the field  in the non-linear regime  as $F^{-\eta}$ with the exponent $\eta$ ranges from $0.2$ to $0.6$, while Eq. (\ref{eq:tand0}) predicts that $\eta=1$.    In our opinion  this non-universal feature of otherwise ``very'' universal  TLSs  requires attention,  because its interpretation can help to understand  the nature of TLSs, while understanding the nature of TLSs should  help to exterminate them in Josephson junctions, which is necessary to make Josephson junction qubits functional  \cite{Lisenfeld2019ReviewTLSs}.   {\it Therefore, here  we investigate  this problem in detail. We  consider critically the earlier proposed  explanation based on TLS dynamic interaction   \cite{Faoro12TLSInt} and suggest the alternative interpretation.  We put forward the hypothesis of  TLS dipole moment power law distribution $P(p) \propto p^{-3+\eta}$ that explains the anomalous field dependence of absorption. }  When deriving Eq. (\ref{eq:tand0}) we assumed approximately identical dipole moments for all TLSs.  The presence of a substantial fraction of large dipole moments in case of a chosen power law distribution dramatically modifies the non-linear absorption field dependence compared to Eq. (\ref{eq:tand0}). 

Indeed,  for this specific distribution the absorption is determined by the crossover domain of dipole moments $\Omega_{c} \sim  1/\sqrt{T_{1}T_{2}}$, Eq. (\ref{eq:lincr}).  The dipole moment at the crossover is defined as  $p(F) \sim \hbar/(F\sqrt{T_{1}T_{2}})$. Then the loss tangent is determined by the linear response theory expression $P_{0}p(F)^2/\epsilon$,  still valid at the crossover, being multiplied by the fraction of such dipole moment $P(p(F))p(F) \approx p(F)^{-2+\eta}$, which yields the desirable dependence $\tan(\delta) \propto F^{-\eta}$.  

{\it We show that this hypothetical  distribution is consistent both with the direct measurement of TLS dipole moments  \cite{Kevin2023PowWeakDep} and the renormalization group theory  considering TLSs as universal collective excitations formed due to the long-range interaction \cite{YuLeggett88,ab96RG,ab98book}. }  This consideration is reported in Sec. \ref{sec:DipDistr}.  In the previous section \ref{sec:FluctAbs} we examine  a TLS absorption affected by its interaction with fluctuators following the earlier suggestions of Ref. \cite{Faoro12TLSInt}, where the explanation of the anomalous non-linear absorption was suggested based on that interaction.  We do not agree with that explanation.  The causes of our disagreement are  briefly explained below, while the detailed consideration is reported in Sec.   \ref{sec:FluctAbs}.

The interaction based   explanation of anomalous absorption was  proposed  in Ref. \cite{Faoro12TLSInt}  for the non-linear regime $\Omega \sqrt{T_{1}T_{2}} \gg 1$  (cf.  Eq. (\ref{eq:lincr})) assuming  the dominating  contribution to the loss tangent  of some resonant TLSs having  specific neighboring fluctuators (see Ref. \cite{Kenyon2001ThermFluct} and Fig. \ref{fig:TLS}).   It  was suggested that if transitions of these   fluctuators shift TLS energy out of resonant domain  during a time $\tau$ less then the inverse Rabi frequency $1/\Omega$, then this resonant TLS absorbs microwaves as in the linear regime in spite of the strong non-linearity.  According to  Ref.  \cite{Faoro12TLSInt}  TLS removal from the resonant domain requires its detuning due to  interaction with fluctuator to exceed  the inverse TLS decoherence rate,   i. e.   $|u|>1/T_{2}$ (see Fig. \ref{fig:TLS}). 
The contribution of such TLSs possessing special neighboring fluctuators  to the loss tangent was estimated as $P_{\rm sp}\tan(\delta_{\rm lin})$, where  $P_{\rm sp}  \approx P_{\rm fl}\ln(1/(\Omega\tau_{\rm min}))$ is the probability that the resonant TLS have  special neighboring fluctuators interacting with TLS stronger than $1/T_{2}$ and possessing   relaxation time less than the inverse Rabi frequency $\Omega$.  Here $\tau_{\rm min}$ is the minimum fluctuator relaxation time that must be sufficiently short $\tau_{\rm min}< \Omega^{-1}\ll T_{1}$,  so we are talking about ``fast'' fluctuators compared to resonant TLSs.   The  logarithmic factor  is originated from the standard assumption of the logarithmically uniform distribution of fluctuator relaxation times above the minimum time.    The predicted  logarithmic field dependence was used to interpret a field dependence of a loss tangent observed experimentally, because a logarithmic dependence is quite similar to the power law dependence with a small exponent \cite{Faoro12TLSInt}.

 The  estimate of the loss tangent in Ref.  \cite{Faoro12TLSInt} raises certain doubts.  It was  assumed there that TLS detuning by fluctuator exceeding $1/T_{2}$ removes it away from the resonant domain.  However, the size of the resonant domain is  equal  to $1/T_{2}$ only in the linear absorption regime (see Fig. \ref{fig:Abs}),  while the nonlinear regime was investigated in Ref.  \cite{Faoro12TLSInt}.   In the non-linear regime $\Omega>\Omega_{c}$,  Eq. (\ref{eq:lincr}), the size of the resonant domain is proportional to the Rabi frequency, Eq. (\ref{eq:tand0}).    Then the probability for resonant TLS  to possess  a neighboring fluctuator capable to remove it away from the resonant domain, i. e.  $u>\delta_{\rm res}$, depends on the field and decreases with increasing the field. For example, if TLS-fluctuator interaction depends on the distance as $\tilde{u}/r^3$ in three dimensions due to elastic or dipolar forces \cite{Hunklinger1986265Rev},  then the probability of a large interaction $\tilde{u}/r^3 > \delta_{\rm res} \approx \Omega \sqrt{T_{1}/T_{2}} $ is proportional to the volume of the space $r^3 = \sqrt{T_{2}/T_{1}} \tilde{u}/\Omega $, where the neighboring fluctuator must be located to remove TLS from resonance. Therefore, the probability to find TLS having  such neighboring fluctuator acquires an additional inverse field dependence $1/F$, since $\Omega \propto F$.  This dependence should enter through the factor $\lambda$  in the loss tangent estimates  reported  in Ref.  \cite{Faoro12TLSInt}.  Consequently, the associated contribution to the loss tangent decreases with the field faster than that in the lack of fluctuators Eq. (\ref{eq:tand0}) due to the logarithmic factor  $\ln(1/(\Omega\tau_{\rm min}))$ and the inverse field dependence derived above.    This is the primarily reason, why we cannot agree with the interpretation proposed in Ref. \cite{Faoro12TLSInt}.   
 
 The actual field dependence of the loss tangent due to contribution of neighboring fluctuators bringing TLS to the linear regime as suggested  in Ref.  \cite{Faoro12TLSInt} is even stronger, as we show in Sec. \ref{sec:FluctAbs}.      We found that the fast fluctuator  contribution to the loss tangent  is due to  representative neighboring  fluctuators and  it is characterized by the field dependence $F^{-4/3}$ that is  intermediate between the dependencies  $F^{-1}$, Eq. (\ref{eq:tand0}),  occurring  in the absence of fluctuators and $F^{-2}$ that takes place  due to  slow fluctuators $\tau_{\rm min} > T_{1}$  \cite{Laikhtman86SpDiffAbs,Galperin88AttenuatSpDiff,ab18nonlinmw}.  This dependence is subject for  experimental verification.  

The paper is organized as following. In Sec. \ref{sec:FluctAbs}  the absorption is considered in the presence of fast fluctuators.  The  model of distributed TLS dipole moments is proposed  and justified in Sec. \ref{sec:DipDistr}.   The summary of the results is presented in Conclusion,  Sec. \ref{sec:Concl}, where we also discuss experimental verification of the theory and  the further  theoretical development.  Appendix contains the derivation of analytical formulas for the fluctuator induced TLS decoherence rate and the  contribution of many fluctuators to the TLS decoherence rate,  all  used in Sec.  \ref{sec:FluctAbs}.

\section{TLS absorption in the presence of fluctuators} 
\label{sec:FluctAbs}

Here we investigate  microwave absorption by TLSs  interacting with  fast fluctuators characterized by the minimum relaxation time $\tau_{\rm min}$ shorter  than that of resonant TLSs ($T_{1}$) following the suggestion of Ref.  \cite{Faoro12TLSInt}  that the minimum fluctuator relaxation time $\tau_{\rm min}$ is shorter than  the  inverse Rabi frequency,  $\Omega^{-1}$, that must be  less than  the TLS relaxation time $T_{1}$ in the non-linear regime of interest opposite to Eq. (\ref{eq:lincr}).  Before we begin the consideration of this regime two  comments are in order about the relevance of the assumption of fast fluctuators and their two-level nature.

In our opinion the assumption $\tau_{\rm min} < T_{1}$ is questionable because in the vast majority  of experiments the temperature is so small ( $T<0.25$K) that the associated thermal energy $E_{T}=k_{B}T$ representing a typical fluctuator energy is smaller than  the resonant TLS   energy $\hbar\omega$ for the typical microwave frequency of $\omega=2\pi\cdot 5$ GHz.    The energy of TLSs serving as  fluctuators should be around the thermal energy to enable them  to switch back and forth between their ground and excited states.  Since TLS relaxation time increases with decreasing its energy $E$ as $E^{-3}$ \cite{Jackle1972}  the minimum fluctuator relaxation time $\tau_{\rm min}$ should be  longer compared to that of resonant TLSs.  

The latter case $T_{1} < \tau_{\rm min}$ was considered in earlier work \cite{Laikhtman86SpDiffAbs,Galperin88AttenuatSpDiff}  and it was treated rigorously in  Ref.  \cite{ab18nonlinmw},  where it was shown  that  fluctuators strengthen the field dependence of the loss tangent in the non-linear regime to $F^{-2}$ compared to the inverse field dependence $F^{-1}$ in the absence of fluctuators \cite{Hunklinger1986265Rev}.    This behavior cannot explain the  observed weakening of the loss tangent field dependence \cite{Siegel2010LossWeakPowDepAl,
Osborn2011NonlinWeakPowDep,Welander2011TLSAbsNLin,Faoro12TLSInt, ab17Katz,Kevin2023PowWeakDep}.   

The regime $\tau_{\rm min}< T_{1}$ considered in Ref. \cite{Faoro12TLSInt} can be relevant   for fluctuators of a different nature than TLSs, so that they still relax faster compared to TLSs.  A weakening of loss tangent field dependence was observed down to $10$mK.  The relaxation time of TLS having similar nature as resonant TLSs at the corresponding thermal energy should be around ten thousand times longer than the time $T_{1}$ for resonant TLSs.  Consequently,  the relaxation time of TLSs,  having the  same energy $E_{T}$ as fluctuators  should be more than ten thousand times longer than the fluctuator relaxation time.  Of course nothing  cannot be excluded but this is very  unforeseen situation.  We still consider this regime because it cannot be excluded and it is also relevant for the high temperatures $\hbar\omega < k_{B}T$ \cite{Galperin88AttenuatSpDiff}. 

Our consideration is limited to two level fluctuators although multilevel fluctuators  can be formed due to, for instance, the nuclear quadrupole interactions  of atoms participating in TLSs \cite{ab06prlNuclQuadr}. To the best of our knowledge there is no observations of more than two state telegraph noise, so we ignore this opportunity.  We are not sure whether two level or multilevel fluctuators were  treated  in Ref. \cite{Faoro12TLSInt}, since it was  assumed there that the fluctuator transition emerging during its  relaxation time $\tau$  removes TLS from  resonance for the time comparable to the TLS  relaxation time $T_{1}\gg \tau$.  Indeed,   after the  next time frame   $\tau$ this two level fluctuator will inevitably come back to its initial state bringing TLS back to resonance much faster than during the time $T_{1}$.   Other neighboring fluctuators can be discarded  during the time of order of $\tau$ under the assumption of their   logarithmically uniform distribution of relaxation times \cite{Faoro12TLSInt}, since  under this assumption the typical relaxation time of the next significant fluctuator (i. e. removing TLS from its resonance) should be much longer compared to $\tau$.  The multilevel nature of fluctuators will substantially reduce the probability of TLS to return back to resonance.  Yet, even if we assume multilevel fluctuators, the field dependent size of the resonant domain in the non-linear regime under consideration should substantially strengthen the field dependence of the loss tangent as it was explained in the end of Sec. \ref{sec:Intr}. 

Following the suggestions of Ref. \cite{Faoro12TLSInt} we  investigate the possibility that TLS - fluctuator interaction can weaken the loss tangent field dependence.  First in Sec.  \ref{subsec:FlMod} we formulate the model of TLS-fluctuator interaction.   Then in Sec.  \ref{subsubsec:IntStrong} we investigate the regime of Ref.  \cite{Faoro12TLSInt},  where neighboring fluctuator of fluctuators bring resonant TLSs into the linear absorption regime.  We show that the probability of such configurations of neighboring fluctuators decreases with the field as inverse squared field in contrast with the experimental observations of substantially weaker dependence.  Then in Sec. \ref{subsubsec:IntWeak} we examine the TLS  absorption affected by the neighboring fluctuators  in their  typical locations  and show that such fluctuators weaken the field dependence of absorption compared to that for  slow fluctuators. However, the predicted field   dependence is stronger than  in the absence of fluctuators.

%We mostly  assume that  a TLS-fluctuator interaction decreases with the distance $r$ as $r^{-3}$, which can be caused by dipole-dipole or elastic interactions in three-dimensions.  In two dimensional materials (e. g.  surface or interface layers)  this interaction can switch  to $1/r^2$ at shorter distances (yet exceeding the thickness of the layer),  see Ref. \cite{smythe1950static} for the electric field and similar behavior is expected for the elastic field. At  longer distances it behaves again as $1/r^3$.  If relevant fluctuators interact with TLSs as   $1/r^2$ in two dimensions, then the results remain qualitatively the same.  However, they can be  different for $1/r^3$ interaction in two dimensions although we found only minor difference as discussed below.   

\subsection{Model of resonant TLS interacting with fluctuators}
\label{subsec:FlMod}  

The resonant TLS interacting with the microwave field and fluctuators  is characterized by the rotating-frame Hamiltonian \cite{Hunklinger1986265Rev} 
\begin{eqnarray}
\widehat{H}= \hbar\left(-\delta S^{z}-\sum_{i}u_{i} \sigma_{i} S^{z}\right) + \hbar\Omega S^{x},  
\label{eq:HTLS}
\end{eqnarray}
where a TLS is denoted by a spin $1/2$ operator $\widehat{S}$,   $\delta$ is the TLS detuning, $u_{i}$ stands for the TLS detuning   induced by the $i^{th}$ fluctuator transitions (i. e.  the TLS-fluctuator interaction, expressed in frequency units). Each fluctuator is characterized by two states $\sigma_{i}=\pm 1/2$ and it randomly transfers   between them with the quasi-period  $\tau_{i}$, expressing fluctuator  relaxation time.  Off-diagonal interactions associated with fluctuators are neglected because they are smaller than diagonal ones by the ratio of the interaction $\hbar u$ and  the fluctuator energy $k_{B}T$.  

Fluctuators are characterized by the probability density determining their distribution in space and with respect to their  relaxation times in the form (cf. Ref. \cite{Faoro12TLSInt})
\begin{eqnarray}
P_{\rm fl}(\tau) = \frac{n_{\rm fl}}{\tau}\theta(\tau - \tau_{\rm min}), 
\label{eq:fluct}
\end{eqnarray}
where $\theta(x)$ is the Heaviside step function,  $\theta(x)=1$ for $x > 0$ and  $0$ for $x<0$,  and  $\tau_{\rm min}$ is the minimum fluctuator relaxation time.   The logarithmically uniform distribution of fluctuator relaxation times is a natural consequence of their exponential sensitivity to the tunneling barrier   separating two fluctuator states \cite{AHV,Ph}.  

We assume that fluctuators have approximately identical probability to occupy both of  their  states.  This significantly simplifies the consideration but does not affect the results at the qualitative level  \cite{Galperin2006TelegrQubitDec,Galperin1988Telegr}. 

%The TLS fluctuator interaction $u$ expressed in frequency units depends on the distance $r$ between TLS and fluctuator.  The spatial density parameter $n_{\rm fl}$ introduces characteristic TLS-fluctuator distance $n^{-1/d}$ where $d$ stands for a space dimension. The interaction at that distance 
%\begin{eqnarray}
%u_{T} \sim u(n^{-1/d})
%\label{eq:uTInt}
%\end{eqnarray}
%stands for a typical TLS interaction with fluctuators having relaxation times of order of $\tau$ provided that $\tau>\tau_{\rm min}$ (here $d$ is a characteristic  dimension).  For instance the density of fluctuators with relaxation times between $\tau$ and $2\tau$ is given by the integral of the probability density function Eq. (\ref{eq:Flpdf}) over relaxation times between $\tau$ and $2\tau$, which yields $n\ln(2) \sim n$. 

%Following Ref.  \cite{Faoro12TLSInt} we consider fluctuators interactions  with TLS exceeding the typical  interaction  $u_{T}$, which are needed to enhance the absorption. 
We characterize  fluctuators using  the  probability density function $F(u, \tau)$ using  TLS-fluctuator interaction $u$ expressed in frequency units.   This function is determined by  the TLS-fluctuator  interaction $U(\mathbf{r})=\hbar u(\mathbf{r})$ and it is defined using  Eq. (\ref{eq:fluct})  as 
\begin{eqnarray}
F(u, \tau)= \frac{n_{\rm fl}\theta(\tau-\tau_{\rm min})}{\tau}\int d^{d}\mathbf{r} \delta(u-u(\mathbf{r})). 
\label{eq:Flpdf}
\end{eqnarray}
%This is valid for the interaction with the closest fluctuator if the  interaction $u$ is much larger  than the TLS fluctuator interaction at the average distance $u_{T}$. 
We evaluate the function $F$ for dimensions $d=3$ and $d=2$, since in some materials TLSs can be located mostly in the surface. In three dimensions we assume that the interaction is dipole-dipole or elastic so its distance dependence takes the form \cite{Hunklinger1986265Rev,YuLeggett88,ab98book}
  \begin{eqnarray}
 u(\mathbf{r})= \frac{\tilde{u}(\mathbf{n})}{r^3}, ~ \mathbf{n}=\frac{\mathbf{r}}{r},
\label{eq:TLSflInt}
\end{eqnarray}
where $\tilde{u}(\mathbf{n})$ is the angular dependent interaction constant \cite{ab98book}. 
We evaluate the integral in Eq.  (\ref{eq:Flpdf})  for this specific  interaction, which yields in three dimensions 
\begin{eqnarray}
F(u, \tau)= \theta(\tau-\tau_{\rm min})\frac{u_{T}}{\tau u^2},  ~ u_{T}=\frac{2\pi}{3}\frac{\tilde{u}_{0}n}{\hbar},
\label{eq:pdf3d}
\end{eqnarray}
where $\tilde{u}_{0}$ is the angular averaged absolute value of the TLS-fluctuator interaction constant  $\tilde{u}(\mathbf{n})$ (cf. Ref. \cite{ab98book}),  and $u_{T}$ is the typical TLS-fluctuator interaction at the average distance $n^{-1/3}$ between them.  If TLSs and fluctuators are located within the thin layer of the thickness $t$ and relevant TLS-fluctuator separation in space exceeds that thickness, then at shorter distances $r$ the interaction behaves as  $\frac{u_{1}(\mathbf{n})}{tr^2}$, due to possible field confinement  inside the layer or as  $\frac{u_{2}(\mathbf{n})}{r^3}$ at longer distances.  In the first case the probability density function is still given by Eq.  (\ref{eq:pdf3d}) with slightly different typical interaction $u_{T}$ that does not make a qualitative difference with the previously considered three dimensional case expressed by  Eq. (\ref{eq:pdf3d}). In the second case we get 
\begin{eqnarray}
F_{2}(u, \tau)= \theta(\tau-\tau_{\rm min})\frac{u_{2T}^{2/3}}{\tau u^{5/3}},  ~ u_{2T}= \frac{(\pi/3)^{3/2}\tilde{u}_{0}(nt)^{3/2}}{\hbar}, 
\label{eq:pdf2d}
\end{eqnarray}
 with $\tilde{u}_{0} \approx \left(<|\tilde{u}(\mathbf{n})|^{2/3}>\right)^{3/2}$.  %We use both Eq. (\ref{eq:pdf3d}) and Eq. (\ref{eq:pdf2d})  to evaluate the fluctuator contribution to the absorption. The virial correction to the loss tangent is evaluated  for three dimensions and then the results are given for two dimensions since their derivation is similar to $3D$. 

The absorption by the resonant TLS is defined by the joint TLS-fluctuator density matrix. This density matrix is   determined  by the stationary solution of generalized Bloch equations \cite{Hunklinger1986265Rev,Carruzzo1994,ab98book}. Below we define such equations for a single neighboring fluctuator, which is sufficient for most of our goals. The generalization to more fluctuators is straightforward. 

We express the density matrix in the form $\rho_{ab}^{cd}$ where two bottom indices $a, b = g$ or $e$ stand for  the ground or excited TLS states,   and two top indices $g$ or $e$ enumerate  ground or excited states of the fluctuator.  For the fluctuator only diagonal elements ($c=d$) are important because of the large energy splitting  ($k_{B}T$) of its two states compared to all other essential energies,  which makes the associated off-diagonal element negligible  \cite{Galperin1988Telegr}.    Then it is convenient to express the density matrix elements in terms of the Bloch vector  projections $S^{x}_{d}= (\rho_{ge}^{dd}+\rho_{eg}^{dd})/2$,  $S^{y}_{d}= i(\rho_{ge}^{dd}-\rho_{eg}^{dd})/2$ and $S^{z}_{d}= (\rho_{gg}^{dd}-\rho_{ee}^{dd})/2$ with the subscript $d=e$ or $g$ characterizing the fluctuator state.  
 
%The state of fluctuator affects the TLS energy $E$ which changes by the characteristic interaction $\hbar u$ during the fluctuator transition between two states. Consequently,  TLS detuning from resonance can be expressed as $\delta+\Delta s^z$ where $s^{z}=\mp 1/2$.  

TLS interaction with the environment (phonons) is characterized by the relaxation time $T_{1}$ and the decoherence time  $T_{2}=2T_{1}$ \cite{ab13LZTh}, while the fluctuator interaction with the environment   is characterized by its relaxation time $\tau$.  
The Bloch vector   projections satisfy   the Bloch equations (index $a$ stands for $g$ or $e$ and complementary index $b$ is $e$ or $g$ states of the fluctuator, respectively,  see  Fig. \ref{fig:TLS}),   
\begin{eqnarray}
\frac{dS^{x}_{a}}{dt}=-\delta S^{y}_{a} + u \sigma_{a} S_{a}^{y} -\frac{S^{x}_{a}}{T_{2}}-\frac{S^{x}_{a}-S^{x}_{b}}{\tau},  
\nonumber\\
\frac{dS^{y}_{a}}{dt}=\delta S^{x}_{a}- u  \sigma_{a}  S_{a}^{x}-\frac{S_{a}^{y}}{T_{2}}+\Omega S_{a}^{z}-\frac{S^{y}_{a}-S^{y}_{b}}{\tau},
\nonumber\\
\frac{dS^{z}_{a}}{dt}=-\Omega S^{y}_{a}-\frac{S^{z}_{a}-1/4}{T_{1}}-\frac{S^{z}_{a}-S^{z}_{b}}{\tau}. 
\label{eq:BlochBas}
\end{eqnarray} 
Eq. (\ref{eq:BlochBas}) differs from the standard Bloch equations \cite{Carruzzo1994}  by the equilibration terms for density matrix in two fluctuator states all inversely proportional to the fluctuator relaxation time $\tau$.  Eq.  (\ref{eq:BlochBas}) is equivalent to the telegraph noise model of  Refs.  \cite{Galperin2006TelegrQubitDec,Galperin1988Telegr} used to  characterize  the influence of fluctuators on TLS spectroscopy.      In a  general case of $N$ fluctuators there are $2^{N}$ possible states  $|a>$ of a subsystem of fluctuators and similar equations should be written for Bloch vectors  $S^{\mu}_{a}$  for each specific state $|a>$. 

%Fluctuators modify TLS loss tangent for each specific configuration of them $c$ around the TLS.   The contribution of each TLS to the loss tangent  is defined by the stationary solutions $S^{y}_{a}$ of Eq.  (\ref{eq:BlochBas}) as    $\delta \tan(\delta)= \hbar\omega\Omega \sum_{a}\langle  S^{y}_{a}\rangle'_{\rm fl} $  \cite{Carruzzo1994,ab98book}.  The change of loss tangent due to fluctuators is defined by this contribution  averaged  over TLS detuning $\delta$ with the weight $P_{0}$ Eq. (\ref{eq:AHWP}) and over all fluctuator configurations, where this contribution substantially exceeds that in the absence of fluctuators ($\tan(\delta_{\rm nl})$, Eq. (\ref{eq:tand0})) since we are interested in the regime where neighboring fluctuators substantially enhance microwave absorption by TLSs.  

The regime considered in Ref.  \cite{Faoro12TLSInt} is realized if the fluctuator configuration around the resonant TLS makes it absorbing as in the linear regime.  The contribution of such special realizations takes the form 
\begin{eqnarray}
\tan(\delta_{\rm sp}) \sim \tan(\delta_{\rm lin})P_{\rm sp}, 
\label{eq:LaraAns}
\end{eqnarray}
where $P_{\rm sp}$ is the probability to find such realization.  In Ref.  \cite{Faoro12TLSInt} the authors argued that this probability decreases logarithmically with the field (Rabi frequency) that can explain the observed loss tangent behavior.  Below in Sec.  \ref{subsubsec:IntStrong} we estimate this probability and find in accord with our arguments in the introduction that it decreases with the field faster than $F^{-1}$.   Consequently,  such configurations cannot explain the observed loss tangent field dependence.  In Sec.  \ref{subsubsec:IntWeak} we examine the  absorption affected by neighboring fast  fluctuators and show that they can weaken  a  loss tangent  field dependence compared to the previously predicted behavior $F^{-2}$  for slow fluctuators. Yet this dependence is stronger than $F^{-1}$ in the absence of fluctuators.

\subsection{Fluctuator configurations   restoring the linear regime}
\label{subsubsec:IntStrong}

We consider a high field regime $\Omega T_{1} > 1$, where the loss tangent is substantially reduced compared to its maximum value given by the linear response theory Eq.  (\ref{eq:tand0}) in the absence of fluctuators.  In this regime  the resonant domain size is determined  by the Rabi frequency,  see Eq. (\ref{eq:tand0}) for $T_{2} =2T_{1}$.    Fluctuators  enhance absorption by means of shifting TLSs away from resonance.  These shifts of TLS resonant frequency  results in its spectral diffusion.  During the absorption time $T_{1}$ the spectral diffusion covers the frequency domain \cite{Galperin88AttenuatSpDiff}  
\begin{eqnarray}
W_{\rm sd} \approx u_{T}
\begin{cases}
    \ln(T_{1}u_{T}), ~ \tau_{\rm min}<\frac{1}{u_{T}},  
    \\   
     \ln\left(\frac{T_{1}}{\tau_{\rm min}}\right), ~ \frac{1}{u_{T}} <\tau_{\rm min}< T_{1},    \\
    \frac{T_{1}}{\tau_{\rm min}},~ T_{1} < \tau_{\rm min}.\\
  \end{cases}   
\label{eq:SDWidth}
\end{eqnarray}

If fluctuators are slow ($T_{1} < \tau_{\rm min}$), the resonant domain is determined by the maximum of the spectral diffusion width or the Rabi frequency as it is shown rigorously in Ref.  \cite{ab18nonlinmw}.  If the  spectral diffusion width exceeds the  inverse TLS relaxation time ($W_{\rm sd} > 1/T_{1}$) the absorption emerges in the linear regime until the saturation  at $\Omega \sim \sqrt{W_{\rm sd} /T_{1}}$.  At larger Rabi frequencies the loss tangent behavior can be approximated by Eq. (\ref{eq:AbsLT}) as 
\begin{eqnarray}
\tan(\delta) \approx \tan(\delta_{\rm lin}) \frac{W_{\rm sd}}{T_{1}\Omega^2},  
\label{eq:SatLT}
\end{eqnarray} 
in a full accord with the rigorous solution of Ref.  \cite{ab18nonlinmw} that does not show an anomalously weak  loss tangent field dependence.   However, the rigorous solution was obtained in the absence of fast fluctuators having relaxation times shorter than  the resonant TLS relaxation time,  i. e.   $\tau_{\rm min} > T_{1}$.   The question then arises, can fast fluctuators  with relaxation times less than $T_{1}$  support  the linear absorption regime with a reasonably high probability?

%This brief answer to this question is "yes" they can by means of raising TLS decoherence rate, but the probability of such  realization of fluctuators decreases with the field too fast  to explain the observed loss tangent behavior. 

To resolve this question we begin with the consideration of  a single fluctuator effect on the absorption.   Let  the fluctuator possessing the relaxation time $\tau$ be coupled with the given TLS by the interaction $u$.  Can it switch the absorption by the given TLS to the linear regime?  Assume  that there is no limitation for the minimum fluctuator relaxation time.  

%Remember that we consider the non-linear regime  suggesting $\Omega^{2} > W_{\rm sd}/T_{1}$, where the width of spectrum $W_{\rm sd}$ induced by the spectral diffusion exceeds the interaction at the average distance $u_{T}$ by at least the logarithmic factor in the regime of interest $\tau_{\rm min}< T_{1}$.  

Fluctuator transitions between two its states  result in decoherence of the TLS.  If the interaction $u$ exceeds the inverse relaxation time $1/\tau$ and TLS detuning $\delta$ is less or comparable with $1/\tau$,  then the fluctuator results in TLS decoherence during the time $\tau$.  This is the straightforward outcome of the Bloch equations Eq. (\ref{eq:BlochBas}).  Indeed,  if for one of the fluctuator  states, e. g. its ground state,  the TLS is in resonance, i. e.  $|\delta +u/2| < 1/\tau$, then for  the other one (excited) the TLS is away from resonance since $|u|>1/\tau$.  Consequently, we can approximately neglect   $x$ and $y$ projections of the Bloch vector   in the state $e$ in Eq.  (\ref{eq:BlochBas}) assuming $S^{x,y}_{g} \gg S_{x,y}^{e}$.   Then  the fluctuator related terms  enters the Bloch  equations  as the additional decoherence channel characterized by the rate $1/\tau$.  The TLS possessing the relaxation time $T_{1}$ and decoherence time $\tau$ absorbs in the linear regime  at sufficiently short  decoherence times $\tau < 1/(\Omega^2 T_{1})$, see Eq. (\ref{eq:lincr}).  The probability to find such neighboring fluctuator is given by the number of fluctuators satisfying the condition $u > 1/\tau > 1/(\Omega^2 T_{1})$.  Using  the probability density function Eq. (\ref{eq:pdf3d}) to evaluate the probability that these inequalities are satisfied we get 
\begin{eqnarray}
P_{\rm sp} \approx \frac{u_{T}}{T_{1}\Omega^2}. 
\label{eq:problinabsfl}
\end{eqnarray}
This probability decreases with the field as an inverse squared field that determines the loss tangent field dependence in Eq.  (\ref{eq:LaraAns}) and thus it is not consistent with  the observed weakening of the loss tangent field dependence.  

The expression for the probability is derived assuming that the TLS - fluctuator interaction $u$ expressed in frequency units exceeds  the inverse fluctuator relaxation time $\tau^{-1}$.   In the opposite limit $u<\tau^{-1}$ the decoherence rate due to fluctuator can be approximated by $1/T_{\rm 2fl} \approx u^2\tau$ (see derivation in  Appendix \ref{App:analsol}  
and Ref.  \cite{Bloembergen1948spindec}  where the similar estimate was obtained for the nuclear spin decoherence rate).  Then the linear absorption takes place under the condition  $u^2\tau > \Omega^2 T_{1}$,  Eq. (\ref{eq:lincr}).  The probability to simultaneously satisfy this condition with the assumed inequality $\tau < 1/u$ evaluated using the probability density function Eq. (\ref{eq:pdf3d})  is also given by Eq. (\ref{eq:problinabsfl}).  Thus a single fluctuator can bring the TLS in the linear absorption regime, but the probability to have such fluctuator decreases with the field very fast, as $F^{-2}$, so such realizations cannot account for the anomalous loss tangent behavior found experimentally.  

One can raise the question, whether more than one fast fluctuator can lead to the desirable behavior.  The probability to have  two neighboring fluctuators contributing to the decoherence rate similarly as required to  ensure the TLS decoherence rate exceeding $\Omega^2 T_{1}$ would be given by $P_{\rm sp}^2$ multiplied by some numerical factor exceeding unity, yet of order of unity  due to the reduction of  requirements for each fluctuator contribution, say, by a factor of $2$ or so.   The field dependence of the probability of such realization is  stronger compared to Eq. (\ref{eq:problinabsfl}). Adding more fluctuators would lead to a stronger field dependence of the associated contribution to the loss-tangent. Consequently,  we do not expect  that the interaction of resonant TLSs with fast fluctuators can explain the observed loss tangent behavior. 

In two dimensional systems with $1/r^{3}$ interaction, the probability density function Eq. (\ref{eq:pdf2d}) should be used to evaluate the probability of the emergence of the linear absorption regime.  The straightforward calculation yields 
\begin{eqnarray}
P_{\rm sp2} \approx \frac{u_{T}^{2/3}}{T_{1}^{2/3}\Omega^{4/3}}. 
\label{eq:problinabsfl2D}
\end{eqnarray}
Although the field dependence here is weaker compared to that in three dimensions it is still stronger than in the non-linear regime without fluctuators.  Thus, it does not explain the anomalous loss tangent filed 
dependence. 

Since the typical relaxation time of fluctuators contributed to the realizations of interest are of order of $1/(\Omega^2 T_{1})$, our consideration is valid only for sufficiently small minimum relaxation time of fluctuators $\tau_{\rm min} < 1/(\Omega^2 T_{1})$.  If the minimum fluctuator relaxation time is longer $\tau_{\rm min} > 1/(\Omega^2 T_{1})$,  then a single fluctuator is not capable to  to bring TLS to the linear absorption regime because the induced decoherence rate cannot exceed $1/\tau_{\rm min}$.  The number of fluctuators of order of $N_{\rm fl} \approx \Omega^2 T_{1}\tau_{\rm min}$ will be needed and the probability for such configuration is exponentially decreasing function of $N_{\rm fl}$. Consequently, such realization  does not contribute to the loss-tangent.

As shown,  we argued that the  interaction based theory cannot explain the anomalous loss tangent field dependence.    We suggest an alternative explanation in Sec. \ref{sec:DipDistr}.   In spite of the non-positive outcome of the present consideration we found that the single neighboring fluctuator can substantially enhance  TLS absorption.  Below in Sec. \ref{subsubsec:IntWeak} we identify  representative fast  fluctuators, which can strengthen absorption. Yet  for that scenario the loss tangent  decreases with the field faster than $1/F$. 

% add graph

\subsection{Absorption due to a ``representative" fluctuator}
\label{subsubsec:IntWeak}

Here we focus on the fast fluctuator effect on the TLS absorption assuming that the main enhancement  is determined by one or few representative  fluctuators located at the average distance from the resonant TLS and possessing certain relaxation time of order of its representative value $\tau_{\rm repr}$.   We assume that  the minimum fluctuator relaxation time $\tau_{\rm min}$ is less than this representative time.  %We also assume that the field is large enough so it exceeds both non-linear thresholds including  Eq. (\ref{eq:lincr}) suggesting $\Omega >1/T_{1}$ (remember that $T_{2}=2T_{1}$) in the absence of fluctuators and the other  one related to the spectral diffusion, i.  e.   $\Omega > \sqrt{W_{T}/T_{1}}$ (see Eq.  (\ref{eq:SDWidth})).  

Let us explain our definition of the representative  fluctuator.  The word ``representative" means that its   contribution to the decoherence rate is comparable to the joint contribution of all other fluctuators.  Consequently, it determines  the size of the resonant domain and,  the loss tangent.  Assume that the interaction $u$ of the representative fluctuator and    the TLS is smaller than its relaxation rate $1/\tau$ (this will be confirmed by the final result). Then, as derived in Appendix \ref{App:analsol} (see also the estimate of a nuclear spin decoherence rate  in Ref. \cite{Bloembergen1948spindec}), the contribution of this fluctuator to the decoherence rate can be expressed as 
\begin{eqnarray}
\frac{1}{T_{\rm 2fl}}=\frac{1}{2\tau}\frac{u^2}{\left(\delta^2+\frac{4}{\tau^2}\right)}. 
\label{eq:DecohRateOptFl}
\end{eqnarray} 
Detuning $\delta$ of absorbing TLSs is of order of  the resonant domain size $\delta_{\rm res}$.  
  
 According to Eq.  (\ref{eq:DecohRateOptFl})  the maximum contribution to the decoherence rate   is provided by fluctuators with the relaxation time of order of inverse detuning  $\delta_{\rm res}$, say, between $\delta_{\rm res}/2$ and $2\delta_{\rm res}$.  There is around one fluctuator possessing the relaxation time close to  the optimum one $1/\delta_{\rm res}$ and having interaction with the TLS  of order  of $u_{T}$, i. e.  located at around the average distance between TLS and fluctuator.  This defines   the representative  fluctuator.  This fluctuator contributes to the decoherence rate as $1/T_{\rm 2fl} \sim u_{T}^2/\delta_{\rm res}$.  The absorption  takes place in the nonlinear regime, since the linear regime emerges only due to fluctuators interacting stronger than the interaction at the average distance  $u_{T}$ (see Sec. \ref{subsubsec:IntStrong}).  Then the resonant domain is determined using Eq. (\ref{eq:tand0})  as $\delta_{\rm res}  \approx \Omega\sqrt{T_{1}/T_{\rm 2fl}}$.    Solving this equation for the resonant domain we calculate it and evaluate the loss tangent using Eq. (\ref{eq:AbsLT}) as 
  \begin{eqnarray}
 \frac{1}{\tau_{\rm repr}} \approx  \delta_{\rm res} = (u_{T}^2\Omega^2T_{1})^{1/3}, ~ \tan(\delta) \approx \tan(\delta_{\rm lin})\frac{u_{T}^{2/3}}{\Omega^{4/3}T_{1}^{2/3}},
  \label{eq:LTOpt}
  \end{eqnarray}
  This estimate based on a single fluctuator is justified for many fluctuators, because the contributions of  fluctuators obey the Levy statistics and their typical contribution is well represented by that of the one located at the average distance and possessing the optimum relaxation time \cite{Raikh00Levy}.  Our analysis of  the effect of all fluctuators reported in Appendix \ref{app:manyfl}  confirms this expectation.  
  
 Eq.  (\ref{eq:LTOpt}) is a subject to  the  three constraints. The resonant domain width, Eq. (\ref{eq:LTOpt}),  must exceed those for the spectral diffusion $W_{T}$, Eq. (\ref{eq:SDWidth}),   and the non-linear absorption without fluctuators, $\Omega$, and the representative fluctuator relaxation time $\tau_{\rm repr}$ should exceed the minimum fluctuator relaxation time $\tau_{\rm min}$.  The first constraint $\delta_{\rm res}>u_{T}$  results in the inequality $u_{T}< 1/\tau$, that was used to derive Eq.   (\ref{eq:DecohRateOptFl}), so it is consistent with the final result expressed by Eq. (\ref{eq:LTOpt}).     
 
If  TLS and fluctuators are of the same origin then the fast fluctuator regime emerges at temperatures where the fluctuator energy $k_{B}T$ exceeds that of resonant TLS $\hbar\omega=k_{B}T_{0}$.   In this regime the loss tangent corresponding to the linear response theory decreases with increasing the temperature as $\tan(\delta_{\rm lin}(T)) \approx  \tan(\delta_{\rm lin})T_{0}/T$,  the resonant TLS relaxation time also decreases as $T_{1}(T) \approx T_{1}T_{0}/T$ and the fluctuator minimum relaxation time decreases as $\tau_{\rm min}(T)\approx T_{1} (T_{0}/T)^3$.   Assuming the interaction $1/R^3$ we also expect the scaling of the TLS-fluctuator interaction in the form   $u_{T}(T)=u_{T}(T_{0})T/T_{0}$  \cite{ab98book}.     Applying all three constraints  formulated in the previous paragraph we found that the regime similar to  Eq. (\ref{eq:LTOpt}) is realized if the TLS-TLS interaction $u_{T}(T_{0})$ at temperature $T_{0}$ exceeds the inverse TLS relaxation time at the same temperature ($u_{T}(T_{0})T_{1}>1$)  and at temperature exceeding $T_{0}(u_{T}(T_{0})T_{1})^2$. The loss tangent and the applicability limits take the form 
\begin{eqnarray}
 \tan(\delta) \approx  \tan(\delta_{\rm lin}(T))\frac{u_{T}(T)^{2/3}}{\Omega^{4/3}T_{1}(T)^{2/3}},
\nonumber\\
 \frac{T}{T_{0}} \sqrt{\frac{u_{T}(T_{0})}{T_{1}}}<\Omega < \frac{T}{T_{0}}u_{T}(T_{0})^2 T_{1}.   
\label{eq:LTOptCond}
\end{eqnarray}

 The predicted behavior $ \tan(\delta) \propto F^{-4/3}$ is intermediate between $F^{-1}$ dependence in the absence of fluctuators and $F^{-2}$ dependence due to slow fluctuators.   
    There are more expected dependencies at intermediate temperatures $T_{0}<T< T_{0}(u_{T}(T_{0})T_{1})^2$.  They will be examined  in a separate work.   Eq. (\ref{eq:LTOpt}) is applicable to two dimensional systems as well, though the additional logarithmic field dependence should be added  there.

\section{Microwave absorption for distributed TLS dipole moments}
\label{sec:DipDistr}

In this section we put forward the hypothesis that the anomalous loss tangent behavior is originated from the TLS dipole moment  power law  distribution $P(d) \propto d^{-3+\eta}$ with $0 \leq \eta <1 $ in a wide domain of dipole moments.  Below in Sec.   \ref{subsec:hyp} we show that  this hypothetical distribution results in  the observed behavior of the loss tangent, i. e.  $\tan(\delta)\propto F^{-\eta}$ or $\ln(1/F)$.   To justify the hypothesis  we show  that  the postulated dipole moment  distribution is consistent with the results of measurements of individual  TLS dipole moments \cite{Kevin2022TLSdata} (see Sec.  \ref{subsec:DistrExp})  and   with the predictions of the renormalization group theory  of collective excitations in the ensemble of two level defects coupled by the long-range interaction $1/R^{d}$ in a $d$-dimensional space \cite{ab96RG,ab96RGpla,ab98book} (see Sec.  \ref{sec:RG}).   Such collective excitations can be formed if TLSs are of electronic nature since electrons possess much larger dipole moments compared to atomic TLSs and, therefore, their  dipole-dipole interaction can be stronger than their elastic interaction. 

\subsection{TLSs with distributed dipole moments: Intensity dependence of absorption} 
\label{subsec:hyp}

Here we calculate the loss-tangent for TLSs with distributed dipole moments. We assume that TLS dipole moments $p$ are distributed between their minimum and maximum values $p_{\rm min}$ and $p_{\rm max}$, respectively, such that $p_{\rm min}\ll p_{\rm max}$ and 
introduce their hypothetical  probability density function  in the simple form 
\begin{eqnarray}
P(\Delta_{0},  \Delta, p) = \frac{P_{0}}{\Delta_{0}}F(p), 
\nonumber\\
F(p)\approx
\begin{cases}
0, ~ p< p_{\rm min}, \\
\frac{p_{\rm min}^{2-\eta}}{(2-\eta)p^{3-\eta}},  ~
p_{\rm min} < p <p_{\rm max},  \\
0, p> p_{\rm max}, 
\end{cases}
~ 0 \leq \eta < 1.    
\label{eq:GenDistr}
\end{eqnarray}
 This hypothetical function discards the domains $p<p_{\rm min}$ and $p>p_{\rm max}$.  It is natural to expect that the function $F(p)$ scales as  $F(p) \propto p^2$ (cf.  Ref.  \cite{ab13echo}) for $p<p_{\rm min}$ and it rapidly (exponentially or so) decreases for $p>p_{\rm max}$.   Under these assumptions the loss tangent intensity dependence is qualitatively correct in various asymptotic regimes considered below with the accuracy to the numerical factor of order of unity, since at high fields the losses will be due to TLSs with $p \sim p_{\rm min}$, while the field independent loss-tangent at small fields is determined by TLSs with dipole moments of comparable to $p_{\rm max}$.

We assume that the temperature is very low so the interaction with thermal TLSs (or fluctuators) can be neglected and the relaxation and dephasing times are connected as $T_{2}=2T_{1}$.  According to the earlier estimate  \cite{ab13LZTh}  the interaction does not affect relaxation and decoherence at temperatures below  $30$mK,  and  the experiments are normally carried out within this temperature domain  \cite{Martinis05,Kevin2023PowWeakDep}.

Assume  that there is no correlations between TLS elastic tensors  and dipole moments and they   are coupled approximately  identically to the elastic field, which is a standard assumption for the TLS model.     Then resonant  TLSs relaxation times are connected to their minimum relaxation time $T_{\rm 1,min}$  as $T_{1}(\Delta_{0})=T_{\rm 1,min}(E/\Delta_{0})^2$   \cite{Jackle1972,ab15TLSnoise}.   Here we use notations different from those in the Introduction and Sec. \ref{sec:FluctAbs}, where the same TLS relaxation time $T_{1}$ was used to estimate the fluctuator contribution to the loss tangent, because the absorption is determined by resonant TLSs having $\Delta_{0} \sim E=\hbar\omega$.  Consequently,  absorbing TLSs possess relaxation times of order of minimum TLS relaxation time denoted there as $T_{1}$.  Since here we perform  accurate calculations,  we assign to each TLS  its own relaxation time $T_{1}$  expressed in terms of the minimum TLS relaxation time  $T_{\rm 1,min}$ as $T_{\rm 1,min}(E/\Delta_{0})^2$.

TLS coupling to the external field is determined  by its Rabi frequency that can be expressed in terms of the maximum Rabi frequency $\Omega_{max}=Fp_{max}$ (remember that $F$ stands for the external microwave  field) as $\Omega =(\Delta_{0}/E_{0})\Omega_{max}\cos(\theta)p/p_{max}$, where $\theta$ is the angle between the field and the TLS dipole moment (remember that for resonant TLSs $E_{0} \approx \hbar\omega$).

  The individual TLS contribution to the loss tangent  takes the form \cite{PhillipsReview}
$$\frac{\Omega^2 T_{2}}{2\epsilon\epsilon_{0}F^2((E/\hbar-\omega)^2T_{2}^2+1+\Omega^2T_{1}T_{2})}.$$ 
This result uses the  stationary  solution of the Bloch equations  $\Omega \langle S^{y}\rangle$,  Eq. (\ref{eq:BlochBas}), determining  the absorption rate  in the absence of fluctuators ($u=0$). 

The loss tangent is defined by the average sum of all TLS contributions within the unit volume, which is equivalent to its averaging over TLS tunneling amplitudes $\Delta_{0}$, angles $\theta$  and TLS dipole moments absolute values $p$  with the probability density function, Eq. (\ref{eq:GenDistr}). The integral can be conveniently expressed as 
\begin{eqnarray}
\tan(\delta)= \frac{4\pi^2 P_{0}p_{\rm min}^2}{(2-\eta)\epsilon R^{\eta}}\int_{R}^{1} \frac{dz}{z^{1-\eta}}\int_{0}^{1}y^2dy\int_{0}^{1}\frac{xdx}{\sqrt{1-x^2}}\frac{1}{\sqrt{1+2x^{-2}y^2z^2\Omega_{\rm max}^2T_{\rm 1,min}^2}},  
\nonumber\\
~ R=p_{\rm min}/p_{\rm max} \ll 1, 
\label{eq:LossTan}
\end{eqnarray}
where  the integration over detuning is already performed and the remaining integration variables are $x=\Delta_{0}/E$,  $y=\cos(\theta)$ and $z=p/p_{max}$.

%\begin{widetext}

%\end{widetext}

There are three distinguishable behaviors of the loss tangent depending on the   product of the maximum Rabi frequency and the minimum relaxation time.   For very small power, $\Omega_{\rm max}T_{\rm 1,min} \ll 1$ the denominator in the integrand in Eq. (\ref{eq:LossTan}) can be replaced with $1$ and the loss tangent  is intensity independent approaching its maximum ($\tan(\delta_{\rm lin})$, cf.  Eq. (\ref{eq:tand0})).      At intermediate powers $\Omega_{\rm min} <1/T_{\rm 1,min}< \Omega_{\rm max}$ ($\Omega_{\rm min}=\Omega_{\rm max}R$) the main contribution to the absorption is originated from dipole moments  $p\sim p_{\rm nl}$ where the saturation of absorption begins, i. e. $p_{\rm nl}F \approx  \hbar/T_{\rm 1,min}$.    Here the loss tangent is reduced by the volume factor $(p_{\rm nl}/p_{\rm max})^3$ and enhanced by the probability density contribution $(p_{\rm nl}/p_{\rm max})^{-3+\eta}$.  Then the loss tangent depends  on the external field as  $\tan(\delta) \propto p_{\rm nl}^{\eta}\propto F^{-\eta}$.  For $\eta=0$ this dependence becomes logarithmic in the form $\tan(\delta) \propto \ln(\hbar/(T_{1,min}Fp_{\rm min})$, that is exactly the dependence that was targeted in Ref. \cite{Faoro12TLSInt}.  %It is important to notice that the power law dependence is distinguishable from the logarithmic one only for $\eta \ln(1/R)> 1$.   
Finally for $1/T_{\rm 1,min} < \Omega_{\rm max}R$ the inverse field dependence of loss tangent takes place as in the standard TLS model with all identical dipole moments \cite{VONSCHICKFUS1977144}.  

We evaluated integrals in Eq.  (\ref{eq:LossTan}) in all these three limits using the probability density function, Eq. (\ref{eq:GenDistr}),  and setting the lower and upper integration limits over dipole moments ($z$) to zero and infinity respectively,  for intermediate  intensities that is justified in case of  $0<\eta<1$. Then we get
\begin{widetext}
\begin{eqnarray}
\tan(\delta) \approx  \frac{4\pi^2}{3 (2-\eta)}\frac{P_{0}p_{min}^2}{\epsilon}
\begin{cases}
    \frac{1-R^{\eta}}{R^{\eta}\eta}, ~ \Omega_{\rm max}<\frac{1}{T_{\rm 1,min}},  
    \\   
     \frac{3}{4(3-\eta)}\frac{\Gamma(\eta/2)\Gamma((1-\eta)/2)\Gamma(1+\eta/2)}{\Gamma((3+\eta)/2)} \frac{1}{(\Omega_{\rm min}T_{\rm 1,min})^{\eta}}, ~ \Omega_{\rm min} <\frac{1}{T_{\rm 1,min}}< \Omega_{\rm max},    \\
    \frac{3\pi}{8\sqrt{2}(1-\eta)}\frac{1}{\Omega_{\rm min}T_{\rm 1,min}},  ~ \frac{1}{T_{\rm 1,min}} < \Omega_{\rm min},\\
  \end{cases}.    
\label{eq:LTLims}
\end{eqnarray}
\end{widetext}
where $\Omega_{\rm min}=R\Omega_{\rm max}$ is the typical Rabi frequency for the minimum dipole moment. 
In the special case of interest of $\eta=0$  we get 
\begin{eqnarray}
\tan(\delta) \approx  \frac{2\pi^2}{3}\frac{P_{0}p_{min}^2}{\epsilon}
\begin{cases}
    \ln(R), ~ \Omega_{\rm max}<\frac{1}{T_{\rm 1,min}},  
    \\   
      \ln\left(\frac{1}{\Omega_{\rm min}T_{\rm 1,min}}\right), ~ \Omega_{\rm min} <\frac{1}{T_{\rm 1,min}}< \Omega_{\rm max},    \\
    \frac{3\pi}{8\sqrt{2}}\frac{1}{\Omega_{\rm min}T_{\rm 1,min}},  ~ \frac{1}{T_{\rm 1,min}} < \Omega_{\rm min},\\
  \end{cases}.    
\label{eq:LTLog}
\end{eqnarray}
The results for small exponents  $\eta$,  i.  e.  $\eta < 1/|\ln(\Omega_{\rm min}T_{\rm 1,min})|$,  given by Eq. (\ref{eq:LTLims}), are formally identical to the results in Eq. (\ref{eq:LTLog}) for $\eta=0$.  Since both the analysis of experimental data reported in Sec.  \ref{subsec:DistrExp}  and the expectations based on the model of TLS formation due to the long-range interactions reported in Sec.  \ref{sec:RG} suggest a nearly vanishing exponent $\eta$ in the power law probability density function of TLS dipole moments, Eq. (\ref{eq:GenDistr}),  here we focus on the regime $\eta=0$. We show that the loss tangent field dependence in this regime  is consistent with the experimental results of Ref.  \cite{Kevin2023PowWeakDep} and,  hopefully, with other observations of similar loss tangent behaviors.

\begin{figure}[h!]
\centering
\includegraphics[width=\columnwidth]{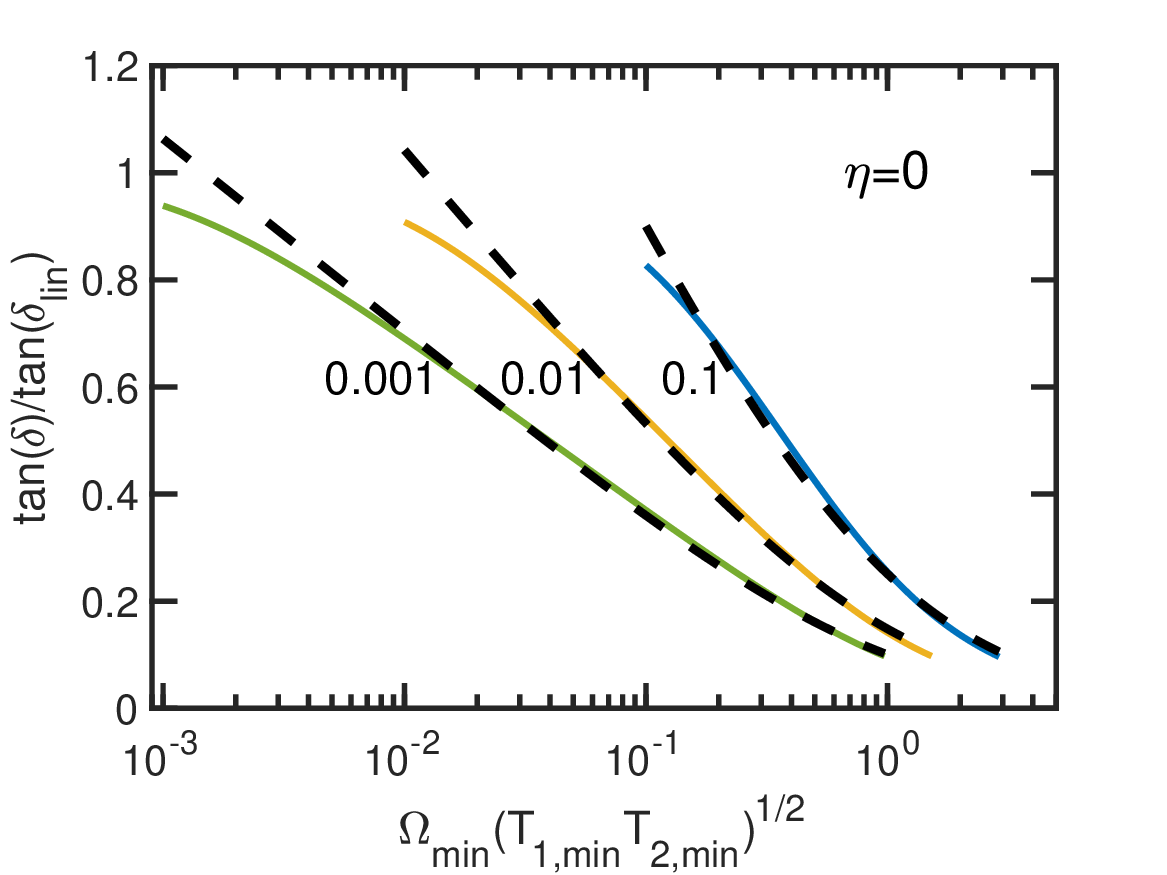}
\caption{\small Loss tangent dependence on the intensity for the TLS dipole moments distributed as $F(p)\propto p^{-3}$ for $\eta=0$.  Ratio $R$ of minimum to maximum dipole moments is shown on the left side of  each graph.  Dashed lines show the fits by the Rabi frequency dependence in the form $\ln(1+C/(\Omega_{\rm min}\sqrt{T_{\rm 1,min}T_{\rm 2,min}}))$ ($T_{\rm 2,min}=2T_{\rm 1,min}$) used in Ref.  \cite{Kevin2023PowWeakDep}.   The fitting parameters $C$ are all equal  $1.05$ independent of the  ratio $R$. }
\label{fig:LTDipDistrLog}
\end{figure}

Indeed, experimental data were fitted  in Ref.  \cite{Kevin2023PowWeakDep} using  the loss tangent field dependence in the form $\tan(\delta) \propto \ln(\gamma/\Omega_{R}+C_{1})$ where $\gamma$ is some constant parameter having  a frequency dimension,  $\Omega_{R}$ is the typical TLS Rabi frequency proportional to the AC field amplitude and $C_{3}\geq 1$ is a constant including the field independent loss of an unknown nature.  This loss can be separated from the field dependent contribution to the loss tangent as $\ln(\gamma/\Omega_{R}+C_{1})=\ln(C_{1})+\ln(\gamma_{1}/\Omega_{R}+1)$, where $\gamma_{1}=\gamma/C_{1}$.  Our consideration does not include field independent loss, while the field dependence expressed by the logarithmic factor $\ln(\gamma_{1}/\Omega_{R}+1)$ is asymptotically  consistent with the predictions of the theory, Eq.  (\ref{eq:LTLog}) obtained using $\gamma_{1}\approx 1/T_{1}$,  and $\Omega_{\rm min}$ instead of $\Omega_{R}$.  
 Indeed, we get a logarithmic field dependence $-\ln(\Omega_{\rm min}T_{1})$ at intermediate  fields $\Omega_{\rm min} <1/T_{1}<\Omega_{\rm max} $ and inverse field  dependence $1/(\Omega_{\rm min}T_{1})$ at large fields  $\Omega_{\rm min}>1/T_{1}$,  Eq. (\ref{eq:LTLog}).     Moreover, we can almost perfectly fit the theoretical predictions for the loss tangent, Eq.  (\ref{eq:LossTan}),  evaluated for $\eta=0$ using the interpolation formula proposed in Ref.   \cite{Kevin2023PowWeakDep}  as illustrated in Fig. \ref{fig:LTDipDistrLog}.  The discrepancy at very small fields corresponds to the linear regime $\Omega_{\rm max}< 1/T_{1}$ (first line in Eq. (\ref{eq:LTLog})), where the logarithmic interpolation does not work.  This regime is not attained in Ref. \cite{Kevin2023PowWeakDep} at classical fields, because the non-linear regime emerges  there even at AC fields corresponding to one photon per cavity.  Smaller fields cannot be treated classically.  The investigation of a  quantum regime is beyond the scope of the present work. 

The logarithmic interpolation formula was suggested in Ref.  \cite{Kevin2023PowWeakDep} based on the prior dynamic interaction based theory \cite{Faoro12TLSInt}.   It makes the  predictions of the loss tangent field dependence quite similar to those in the present work.  As we argued in the Introduction the dynamic interaction based theory  underestimated the size of the resonant domain within the non-linear regime, where it is determined by a Rabi frequency proportional to the AC field rather than the TLS decoherence rate.  As it is shown in Sec.  \ref{sec:FluctAbs}, if the definition of a resonant domain would be corrected, then the field dependence is not logarithmic, but it is even stronger than the inverse field dependence realized in the absence of fluctuators. Consequently,  we do not believe that the explanation of Ref.  \cite{Faoro12TLSInt} is applicable to the experimental data under consideration.

%Thus, the proposed theory is consistent with the interpolation formula used to fit the experimental data of Ref. \cite{Kevin2023PowWeakDep} within the field domain, where it is technically applicable (for more than one photon per cavity).    Anomalous loss tangent behavior discovered by the other groups is quite similar to the observations of Ref.   \cite{Kevin2023PowWeakDep}.  Thus,  we hope that the proposed theory will be applicable there as well.  We do not attempt to fit any specific experimental  data here, because this requires performing the complicated tasks including a separation of  field independent contributions and an  investigation of a quantum regime.  These considerations are beyond the scope of the present work.}

  \subsection{Comparison of theory with dipole moment probability density function  determined by direct measurements \cite{Kevin2022TLSdata}.
\label{subsec:DistrExp}
 } 
  % Integral num FigLTEta05.eps eta=0.5
  % FigLTEta025.eps
% Renormalization group RGRndN27.eps
% Lisenfeld2017ProbeManyTLS
A typical TLS dipole moment can be determined experimentally using a variety of methods.  For example  two and three pulse dielectric echo  measurements permit us to  extract a typical value of dipole moment and to obtain some knowledge about its distribution in the integrated form \cite{Baier1988DielectricEcho,Enss1996DielEcho,ab13echo}.  TLS dipole moment can also be estimated measuring absorption in the presence of time-varying bias.  However, these methods do not probe  the distribution of dipole moments particularly including its tails as expressed by Eq.  (\ref{eq:GenDistr}),   which are potentially responsible  for the anomalous field dependence of the loss tangent,   considered in the present work.

\begin{figure}[h!]
\centering
\includegraphics[width=\columnwidth]{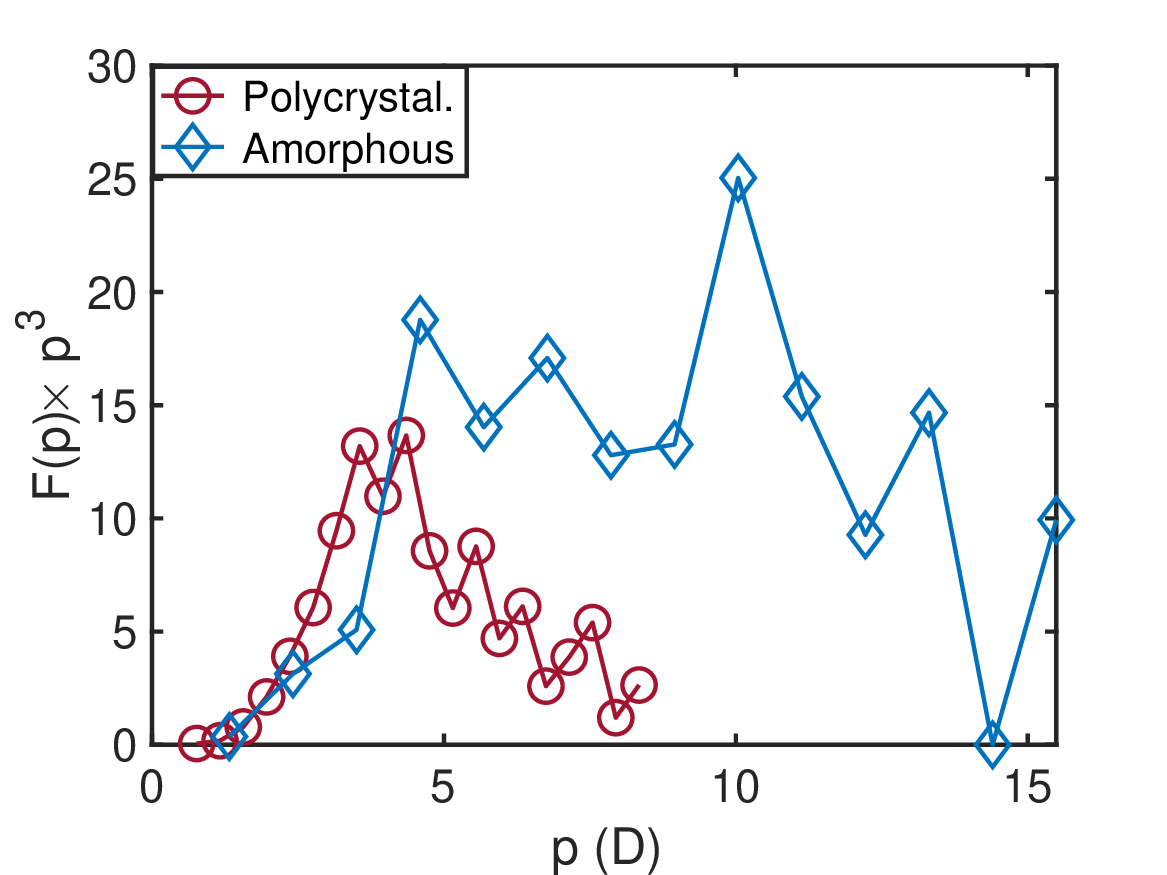}
\caption{\small Dipole probability density function  vs.  dipole moment rescaled by the factor $p^3$ for polycrystalline and amorphous Al oxides \cite{Kevin2022TLSdata}. }
\label{fig:PDFExpFit1}
\end{figure}

To investigate  TLS distribution one should measure  individual TLS dipole moments.   TLS dipole moments were measured using  resonant transmission probes in the presence of the external bias field $\mathbf{F}$ modifying TLS resonant frequency as $-\mathbf{F}\mathbf{p}/\hbar$   \cite{ab16Bahman}.   Measurements of many TLS dipole moments were reported in Refs. \cite{ab16Bahman,Lisenfeld2017ProbeManyTLS,Kevin2022TLSdata} in Al oxide junctions.   The  TLS  dipole moment probability density functions were reported in Ref.  \cite{Kevin2022TLSdata} for poly-crystalline and amorphous aluminum oxide films.  
Here we analyze those data, extracted using the online tool \cite{DataGet} with the permission of the authors of Ref.  \cite{Kevin2022TLSdata}.   One  comment is in order before starting an analysis.  The experimental  data are obtained for the probability density function of dipole moment projections, rather than  dipole moment absolute values considered in Sec.  \ref{subsec:hyp}.   However,   it is straightforward to check that for $p_{\rm min} < p \ll p_{\rm max}$ the probability density function for dipole moment projections possesses the same power law tail as that for TLS dipole moment absolute values.

 The hypothetical  form of the probability density function $F(p)$, Eq. (\ref{eq:GenDistr}), is related to the deep tail of the actual probability distribution.  If at large dipole moments $p>p_{\rm min}$ this function decreases with increasing the dipole moment as  $F(p) \propto 1/p^{3-\eta}$ with $0\leq \eta < 1$, then vast majority of TLSs possess a dipole moment $p \sim p_{\rm min}$.  However,  the TLS contribution to the loss tangent is proportional to their  squared  dipole moments (for instance,  $\tan(\delta) \propto \int dp p^2F(p)$ in the linear regime) and, therefore , it is determined by the tail of $F(p)$ at large dipole moments.  In other words {\it the absorption is due to   a small fraction of TLSs with large dipole moments, while the majority of TLSs contribute negligibly.}
 Thus the main goal of our analysis is to find out whether the TLS probability density function found experimentally in Ref.  \cite{Kevin2022TLSdata} obeys the asymptotic behavior given by Eq. (\ref{eq:GenDistr})  at large dipole moments or not.  
 
 The probability density function was analyzed in Ref.  \cite{Kevin2022TLSdata}. It was fitted by Gaussian and gamma-distributions.  However, this analysis targeted   typical TLS dipole moments and it is not relevant for distribution tails, which are the targets of our consideration.

%figKevinFit2.eps
\begin{figure}[h!]
\centering
\includegraphics[width=\columnwidth]{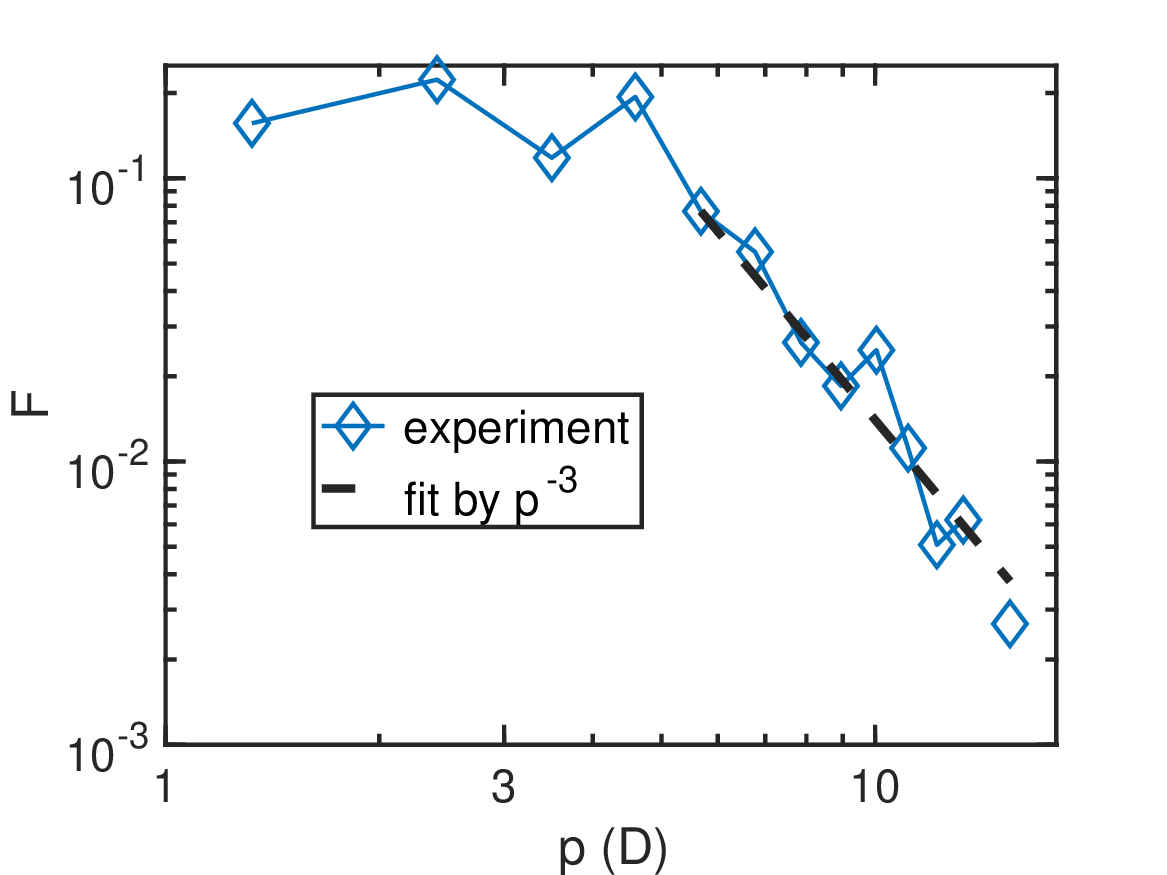}
\caption{\small The probability density function  vs.  dipole moment for amorphous Al oxide \cite{Kevin2022TLSdata} compared with the theoretical expectations.  The point with zero value of the function is omitted, because it cannot be shown in the logarithmic scale.}
\label{fig:PDFExpFit}
\end{figure}

The loss tangent is determined by the tail of the probability density function $F(p)$ of dipole moments  if it decreases with increasing the dipole moment as $p^{-3}$ or slower.   To verify whether this is true or not for the experimental data of Ref.  \cite{Kevin2022TLSdata} we evaluated the product of that function obtained experimentally and the cube of a dipole moment and report the results in Fig. \ref{fig:PDFExpFit1} both for polycrystalline and amorphous aluminum oxide films.  According to the graph the  distribution function tail is not significant for polycrystalline films, where the probability density function decreases with increasing the dipole moment much faster than $p^{-3}$.  However, in the amorphous films the tails are significant and should contribute to the loss tangent.  

We also attempted to fit  the tails of the probability density functions for dipole moment projections using the power law fit.  For the poly-crystalline sample the best power law fit was obtained using a dependence  $F(p) \propto p^{-5}$ (not shown).   However, for the amorphous sample the probability density  function  follows the  $p^{-3}$ law reasonably well  for dipole moments exceeding $4$ D as shown in Fig. \ref{fig:PDFExpFit}, so it is consistent with the proposed hypothetical distribution.  We fit only the tail of the distribution $p>4D$ assuming that the minimum dipole moment $p_{\rm min}$ is of order of few D.  At smaller dipole moments the power law distribution is obviously irrelevant.

Based on the observed behavior  of the probability density functions in polycrystalline and amorphous samples we predict that the anomalously weak dependence of loss tangent on the AC field intensity should be seen in amorphous samples, while the non-linear loss tangent in polycrystalline samples should behave according to the standard TLS model.  The present work should hopefully encourage experimentalists to collect more data and verify the relevance of the present model.   Below we discuss some theoretical grounds behind the power law dipole moment distribution. 

\subsection{Universal probability density function of collective excitations in the presence of  long-range interactions}
\label{sec:RG}
% strong interaction at high energy converts to a very small interaction at low energy (temperature)

The power law dependence of the probability density function  on the dipole moment, $F(p)\propto p^{-3+\eta} $, expressed by   Eq. (\ref{eq:GenDistr}),  emerges naturally  for coupling constants of low energy collective excitations of interacting two-level defects (Ising pseudospins $1/2$) coupled by the long-range interaction $u/r^d$ in $d$-dimensions \cite{ab96RG,ab96RGpla}.   The quoted  theory was developed to interpret the quantitative universality of   TLSs in amorphous solids as those collective excitations,  following the earlier suggestion of Refs.   \cite{YuLeggett88,Yu89}.  The renormalization group analysis of this  model  with random uncorrelated interaction constants $u$ between each pair of defects  \cite{ab96RG,ab96RGpla} predicted that the  low energy  excitations are distinguished by the number of participating defects  $n$ and possess the universal densities of states $P_{n} \propto \ln(n)/n^{2}$.  In that specific model the coupling constant for $n$-defect excitations scales with $n$ as $\gamma \propto \sqrt{n}$. The distribution of coupling constants $\gamma$ behaves consequently as 
\begin{eqnarray}
F(\gamma) \approx \int_{0}^{\infty}dn P_{n}\delta(\gamma-\sqrt{n})   \propto \ln(\gamma)/\gamma^3, 
\label{eq:coupldistr}
\end{eqnarray}
which is similar to the desirable distribution of dipole moments Eq. (\ref{eq:GenDistr}).

The behavior  $P_{n}\propto \ln(n)/n^2$ is connected with the approximate  independence of a  susceptibility $\sum_{n=1}^{\infty}P_{n}n$ on the system size. It remains  non-zero in the limit of zero temperature and an  infinite interaction radius in spite of all vanishing probability densities $P_{n}$ in this limit due to the dipolar gap   \cite{Efros75,ab96RG,ab96RGpla}.  Then the divergence of the sum $\sum_{n=1}^{\infty}P_{n}n$ at large $n$ is required to keep this sum constant in the limit of an infinite interaction radius.  
Similarly for the system with the dipole-dipole interaction   the susceptibility $\int_{0}^{\infty} dp F(p)p^2$ must remain non-zero  in spite of vanishing probability density $F(p)$ due to the dipolar gap.   This  requires  the divergence of the integral  $\int_{0}^{\infty} dp F(p)p^2$ at large $p$ (otherwise, the number of excitations  $\int_{0}^{\infty} dp F(p)$  must grow with increasing the system size, that conflicts with the renormalization group theory  \cite{ab96RG}).  Consequently,  the behavior $F(p) \propto p^{-3+\eta}$ with $\eta \geq 0$ is expected for dipole moments of collective excitations possibly representing TLSs.   %Consequently,  we do expect that the renormalization group theory similar to that developed in Refs.  \cite{ab96RG,ab98book}  can explain the hypothetical distribution of TLS dipole moments Eq. (\ref{eq:GenDistr}). 

The predicted distribution of dipole moments  is relevant if the dipole-dipole interaction is the strongest long-range interaction in the system.  To our knowledge in the vast majority of amorphous dielectrics elastic interaction is stronger than the dipole-dipole interaction \cite{ab95dipgap,ab98book}.  Indeed, TLSs of atomic nature possess very small dipole moments of order or less than few D because heavy atoms can tunnel only to very short distances. However, if TLSs are of electronic nature, then the tunneling distances can be much longer up to nanometers and the associated dipole-dipole interaction can exceed the elastic one.  Consequently, we expect that the materials showing anomalous non-linear absorption possess TLSs of electronic nature formed by e. g. localized quasi-particles \cite{PGustafsson2013PRBElectTLS, Kafanov2008ElTLS}.

\section{Conclusions}
\label{sec:Concl} 
% add about temperature independence at low temperature as a styraightforward verification of interaction irrelevance

We investigate the origin of the anomalous non-linear microwave absorption by two level systems (TLSs) in Josephson junction qubits  characterized by a weak  loss-tangent field dependence $F^{-\eta}$, where $\eta<1$,   compared to the inverse field dependence expected theoretically.    Two possible sources  of this behavior  were considered.  First, we investigated the effect of  interaction of  resonant TLSs with fluctuators  (e. g.  TLSs with energy splitting of order of  the thermal energy $k_{B}T$). We found, in agreement with earlier work \cite{Laikhtman86SpDiffAbs,Galperin88AttenuatSpDiff,ab18nonlinmw},  that   the loss tangent field dependence is strengthened in the presence of interaction  compared to that  in the absence of interaction ($F^{-1}$).  Consequently,  TLS interaction  cannot lead to the experimentally discovered weakening of the loss tangent field dependence.  Yet,  we predicted  a new non-linear absorption regime characterized by the field dependencies $F^{-4/3}$  emerging if fluctuators interacting with  resonant TLSs relax much faster than  resonant TLSs. 
This prediction awaits  experimental verification.   

Second, we attempted to interpret the experiment  assuming a hypothetical power law distribution of TLS dipole moments $p$ in the form $F(p)\propto p^{\eta-3}$. This distribution, indeed, results in the desirable loss tangent  behavior; yet, the proposed hypothesis needs justification.  For a  justification, we find on the one hand that the proposed distribution of dipole moments is consistent with the experimental observations in amorphous aluminum oxide films \cite{Kevin2022TLSdata}. On the other hand,  it is consistent with the expected distribution of coupling strengths (dipole moments in our problem) of low-energy collective excitations in systems with long-range interactions  \cite{ab96RG,ab96RGpla,ab98book}.  {\it Thus the anomalous loss tangent field dependence signals that the TLSs could be,  indeed, represented by such universal excitations.} 
The relevance of theory requires dominating TLS dipole-dipole interactions over  elastic interactions.  This requires large absolute values of TLS dipole moments, which   is a sign of  the electronic nature of TLSs in materials, where the anomalous loss tangent behavior is observed.  Both comparison of the hypothetical dipole moment distribution with the experiment and the theoretical analysis  considering the low energy collective excitations suggest that the power $\eta$ in the dipole moment probability density function is close to zero.  This assumption is consistent with the interpolation formula used in Ref. \cite{Kevin2023PowWeakDep} to interpret experimental data. 

The proposed theory   needs further experimental verifications and theoretical developments.  Although we ruled out the interaction with two level  fluctuators as a source  of anomalous non-linear absorption,  the interaction effect can be different if, for instance,  the fluctuators are multilevel.    However, if the anomalous absorption is due to any dynamic interaction (like the interaction with fluctuators) it should be dramatically sensitive to the  temperature.   Particularly,   it should disappear with decreasing the temperature,   since at zero temperatures all  dynamic excitations must be    frozen out.  Consequently, the investigation of loss tangent temperature dependence at low temperatures can  test the significance of dynamic TLS interactions.  The lack of temperature dependence will rule out any dynamic interaction.   We did not find any  evidences of temperature dependence of a non-linear absorption in the recent work \cite{Kevin2023PowWeakDep}. 

The hypothesis of the  power law distribution of TLS dipole  moments can be verified by simultaneous   measurements   of many TLS dipole moments in a single sample similarly to Ref. \cite{Kevin2022TLSdata}.    The  theory interpreting TLSs as collective excitations formed due to the long-range interaction \cite{ab96RG,ab96RGpla,ab98book}   needs to be generalized  to the dipole-dipole interactions.   The  dipole moment statistics and their correlations with elastic interactions can be then  investigated  within the framework of that theory.   We hope that the local experimental probes of TLSs like the recently developed scanning gate imaging technique \cite{TLSScan2024graaf} will examine a composite nature of TLSs as a source of their power law dipole moment distribution as proposed in the present work.

\begin{acknowledgments}
This work is partially supported by the National Science Foundation (CHE-1462075), the Tulane University Carol Lavin Bernick Faculty Grant and the NSF-Lousiana Board of Regents LINK Program.   I am grateful to  Moshe Schechter for very useful discussions during my visit to Israel in winter 2024 and to Kevin Osborn  for stimulating discussions, suggestions and for sharing the experimental results  for the TLS dipole moment statistics from Ref.  \cite{Kevin2022TLSdata} used to compare the experiment with the theory.     

I acknowledge Organizers and Participants of the Aspen Winter Conference ``Noise and Decoherence in Qubits" for useful and stimulating discussions.    I also acknowledge the Organizers of this extra-ordinary event and   Duke Quantum Center  for supporting my travel to Aspen. 

\end{acknowledgments}

% In-situ scanning gate imaging of individual two-level material defects in live superconducting quantum circuits M. Hegedüs, R. Banerjee, A. Hutcheson, T. Barker, S. Mahashabde, A. V. Danilov, S. E. Kubatkin, V. Antonov, S. E. de Graaf
% Promising scanning measurements - worth to mention
% Oxygen vacancies in niobium pentoxide as a source of two-level system losses in superconducting niobium D. Bafia ,* A. Murthy , A. Grassellino, and A. Romanenko Елецтрониц тыпе оф ТЛС соурце, нотице ит

\bibliography{MBL}

\appendix

\section{Definition of the fluctuator induced decoherence rate}
\label{App:analsol}

Here we derive the expression  for the absorption  in the presence of a single fluctuator to justify our approximation for the decoherence rate Eq. (\ref{eq:DecohRateOptFl}) made in the main text. 

We reexpress the stationary version of Eq.  (\ref{eq:BlochBas}) from the main text for sums and differences of the Bloch vector projections  $S^{\alpha}_{\pm}=S^{\alpha}_{g}\pm S^{\alpha}_{e}$ as 
\begin{eqnarray}
0=-\delta S^{y}_{+} + \frac{u}{2}S^{y}_{-} -\frac{1}{T_{2}}S^{x}_{+},  
\nonumber\\
0=-\delta S^{y}_{-} + \frac{u}{2}S_{y}^{+} -\left(\frac{2}{\tau}+\frac{1}{T_{2}}\right)S^{x}_{-},   
\nonumber\\
0=\delta S^{x}_{+}- \frac{u}{2} S_{-}^{x}-\frac{1}{T_{2}}S_{+}^{y}+\Omega S_{+}^{z}, 
\nonumber\\
0=\delta S^{x}_{-}- \frac{u}{2} S_{+}^{x}-\left(\frac{2}{\tau}+\frac{1}{T_{2}}\right)S_{-}^{y}+\Omega S_{-}^{z}, 
\nonumber\\
\frac{1}{2T_{1}}=\Omega S^{y}_{+}+S^{z}_{+}\frac{1}{T_{1}},
\nonumber\\
0=\Omega S^{y}_{-}+\left(\frac{2}{\tau}+\frac{1}{T_{1}}\right)S^{z}_{-}. 
%\frac{dS^{z}_{a}}{dt}=\Omega S^{y}_{a}+\frac{S^{z}_{a}-1/4}{T_{1}}-\frac{S^{z}_{a}-S^{z}_{b}}{\tau},
\label{eq:Blochpm}
\end{eqnarray} 
Solving the last equation for $S_{-}^{z}$ we get rid of  $S_{-}^{z}$ in the fourth equation reducing the number of variables as 
\begin{eqnarray}
%0=-\delta S^{y}_{+} + \frac{u}{2}S^{y}_{-} -\frac{1}{T_{2}}S^{x}_{+},  
%\nonumber\\
%0=-\delta S^{y}_{-} + \frac{u}{2}S_{y}^{+} -\left(\frac{2}{\tau}+\frac{1}{T_{2}}\right)S^{x}_{-},   
%\nonumber\\
%0=\delta S^{x}_{+}- \frac{u}{2} S_{-}^{x}-\frac{1}{T_{2}}S_{+}^{y}+\Omega S_{+}^{z}, 
%\nonumber\\
0=\delta S^{x}_{-}- \frac{u}{2} S_{+}^{x}-\left(\frac{2}{\tau}+\frac{1}{T_{2}}+\frac{\Omega^2}{\frac{2}{\tau}+\frac{1}{T_{1}}} \right)S_{-}^{y},
%\nonumber\\
%\frac{1}{2T_{1}}=\Omega S^{y}_{+}+S^{z}_{+}\frac{1}{T_{1}}.
\label{eq:Blochpm1}
\end{eqnarray} 
Since in the regime of interest $\tau < T_{1}, T_{2}$, we neglect $1/T_{1}$ and $1/T_{2}$ compared to $1/\tau$ and simplify Eq. (\ref{eq:Blochpm1}) as 
\begin{eqnarray}
0=\delta S^{x}_{-}- \frac{u}{2} S_{+}^{x}-\frac{2}{\tau}\left(1+\frac{\Omega^2\tau^2}{4} \right)S_{-}^{y}.
\label{eq:Blochpm2}
\end{eqnarray} 
The second and fourth equations can be solved for $S^{x,y}_{-}$ pseudospin components  as 
\begin{eqnarray}
S^{x}_{-}=\frac{u}{2}\frac{\delta S^{x}_{+}+\frac{2}{\tau}\left(1+\frac{\Omega^2\tau^2}{4} \right)S^{y}_{+}}{\delta^2+\frac{4}{\tau^2}\left(1+\frac{\Omega^2\tau^2}{4} \right)},
\nonumber\\
S^{y}_{-}=\frac{u}{2}\frac{\delta S^{y}_{+}-\frac{2}{\tau}S^{x}_{+}}{\delta^2+\frac{4}{\tau^2}\left(1+\frac{\Omega^2\tau^2}{4} \right)}.
\label{eq:Blochpm3}
\end{eqnarray} 
Substituting the solutions Eq.  (\ref{eq:Blochpm3}) into first and third equation in Eq. (\ref{eq:Blochpm2}) we left with only three equations in  the form 
\begin{eqnarray}
0=-\delta S^{y}_{+} \left(1-\frac{u^2}{4\left(\delta^2+\frac{4}{\tau^2}\left(1+\frac{\Omega^2\tau^2}{4} \right)\right)}\right) -S^{x}_{+}\left(\frac{u^2}{4}\frac{\frac{2}{\tau}}{\delta^2+\frac{4}{\tau^2}\left(1+\frac{\Omega^2\tau^2}{4} \right)} +\frac{1}{T_{2}}\right),  
\nonumber\\
0=\delta S^{x}_{+} \left(1-\frac{u^2}{4\left(\delta^2+\frac{4}{\tau^2}\left(1+\frac{\Omega^2\tau^2}{4} \right)\right)}\right) - S^{y}_{+}\left(\frac{u^2}{4} \frac{\frac{2}{\tau}\left(1+\frac{\Omega^2\tau^2}{4} \right)}{\delta^2+\frac{4}{\tau^2}\left(1+\frac{\Omega^2\tau^2}{4} \right)}+\frac{1}{T_{2}}\right)+\Omega S_{+}^{z}, 
\nonumber\\
\frac{1}{2T_{1}}=\Omega S^{y}_{+}+\frac{1}{T_{1}}S^{z}_{+}.
\label{eq:Blochpm4}
\end{eqnarray} 
Since,  according to our assumption, $u\tau \ll 1$ we neglect the subtracted term compared to unity in the first parenthesis in the first and second equations.  We also neglect $\Omega^2$ term in denominators as compared to the squared detuning $\delta^2 \sim \delta_{\rm res}^2$ since we are interested in the fluctuator induced major increase  in absorption that emerges at $\delta_{\rm res}> \Omega$.  
Then we get the absorption rate $\Omega S^{y}_{+}$ in the form 
\begin{eqnarray}
\Omega S^{y}_{+} =\frac{\Omega^2\left(\frac{1}{T_{\rm 2fl}}+\frac{1}{T_{2}} \right)}{\delta^2+\Omega^2T_{1}\left(\frac{1}{T_{\rm 2fl}}+\frac{1}{T_{2}} \right) + \left(\frac{1}{T_{\rm 2fl}}+\frac{1}{T_{2}} \right)\left(\frac{\left(1+\frac{\Omega^2\tau^2}{4} \right)}{T_{\rm 2fl}}+\frac{1}{T_{2}} \right)},
\nonumber\\
\frac{1}{T_{\rm 2fl}}=\frac{u^2}{4}\frac{\frac{2}{\tau}}{\delta^2+\frac{4}{\tau^2}}.
%\frac{1}{\Omega^2T_{1}}\left(\frac{\delta^2}{\left(\frac{u^2}{4}\frac{\frac{2}{\tau}}{\delta^2+\frac{4}{\tau^2}\left(1+\frac{\Omega^2\tau^2}{4} \right)} +\frac{1}{T_{2}}\right)}+%\left(\frac{u^2}{4} \frac{\frac{2}{\tau}\left(1+\frac{\Omega^2\tau^2}{4} \right)}{\delta^2+\frac{4}{\tau^2}\left(1+\frac{\Omega^2\tau^2}{4} \right)}+\frac{1}{T_{2}}\right)
%\right) }.
\label{eq:Blochpm5}
\end{eqnarray} 
This form is almost identical to the stationary solution of the ordinary Block equation with relaxation and decoherence rates $1/T_{1}$ and $1/T_{\rm 2fl}+1/T_{2}$, respectively, except for the  third term in the denominator containing extra factor $1+\Omega^2\tau^2$. It turns out that this term can be neglected compared to the first term in the denominator if the contributing fluctuator relaxation time is less than the TLS relaxation time $T_{1}$ (according to the definition of $T_{\rm 2fl}$,   it  always exceeds $\tau$ under the assumption $u \ll \delta \sim \delta_{\rm res}$). This justifies Eq. (\ref{eq:DecohRateOptFl}) in the parametric domain of interest $\tau_{\rm min} < T_{1}$,  where this equation was used in the main text.

\section{Estimate  of contributions of many fluctuators to decoherence rate. } 
\label{app:manyfl}

Here we consider the contribution of neighboring  fluctuators to a resonant TLS decoherence rate in the regime,  where the characteristic  TLS-fluctuator interaction $u_{T}$ (see Eq. (\ref{eq:pdf3d})) is less than the  TLS detuning $\delta$.  In Sec. \ref{subsubsec:IntWeak} we estimated that contribution replacing all fluctuators  with one representative  fluctuator possessing the relaxation time $\tau \approx 1/\delta$  and here we examine  the relevance of this approach. 

In the regime of interest the single fluctuator contributes  to the decoherence rate according to  Eq. (\ref{eq:DecohRateOptFl}).  For  many fluctuators similar consideration  replaces   this contribution with  the algebraic sum of all fluctuator contributions that takes the form 
\begin{eqnarray}
r=\frac{1}{T_{\rm 2fl}}=\sum_{i}\frac{1}{\tau_{i}}\frac{u_{i}^2}{2\left(\delta^2+\frac{4}{\tau_{i}^2}\right)}, 
\label{eq:appDecohRateFl}
\end{eqnarray}   
where index  $i$ enumerates  all neighboring fluctuators.  The decoherence rate $r$ depends on the spatial positions and tunneling amplitudes of all surrounding fluctuators.   We estimate the typical value of this rate evaluating  its distribution function  $P(r)$.   The  Fourier transform of this distribution function reads  \cite{BlackHalperin77,ab98book} ($n$ is the spatial density of fluctuators, see Eq. (\ref{eq:fluct})) 
 \begin{eqnarray}
p(x)=\langle e^{ixr}\rangle = \exp\left(-n\int_{0}^{\infty}\frac{d\tau}{\tau}\int_{-\infty}^{\infty}d\delta\int d\mathbf{r} \left(1-\exp\left[ix\frac{\tilde{u}(\mathbf{n})^2}{2\tau r^6\left(\delta^2+\frac{4}{\tau^2}\right)}\right]\right)  \right).
\label{eq:appDecohpdfFour}
\end{eqnarray} 
For the sake of simplicity we assume that the minimum decoherence time $\tau_{\rm min}$ is very small so it is replaced with $0$. This assumption is justified if the minimum relaxation rate is less than its representative value $1/\delta$, see  Eq. (\ref{eq:DecohRateOptFl}). 

The integral in Eq. (\ref{eq:appDecohpdfFour}) can be evaluated analytically as 
  \begin{eqnarray}
\langle e^{ixr}\rangle = e^{-\frac{1-i\cdot{\rm sign}(x)}{\sqrt{2}}\sqrt{|x|R_{*}}}, ~~ R_{*}=\frac{\Gamma(1/4)^4u_{T}^2}{2|\delta|}. 
\label{eq:appDecohpdfFourRes}
\end{eqnarray} 
The probability density function for different relaxation rates is determined by the inverse Fourier transform of Eq. (\ref{eq:appDecohpdfFourRes}) that takes the form 
  \begin{eqnarray}
  p(r)=\frac{1}{R_{*}}P(r/R_{*}), ~ P(x)=\frac{1}{\pi}{\rm Re}\int_{0}^{\infty}d\xi e^{-i\xi x-\sqrt{\xi}\frac{1-i}{\sqrt{2}}}.
\label{eq:appDecohpdfRes}
\end{eqnarray}
Thus the typical  fluctuator contribution $R_{*}$ to the decoherence rate of TLS having detuning $\delta$  is equivalent  to the contribution of a representative   fluctuator possessing the relaxation time $\tau \sim 1/\delta$ and coupled to the resonant TLS with a strength  $u_{T}$ as it was  suggested in Sec. \ref{subsubsec:IntWeak}.   

\end{document}